\documentclass{emulateapj}
\usepackage{amsmath}
\input{epsf}
\usepackage{graphicx}
\usepackage{epstopdf}
\usepackage{bm}
\usepackage[breaklinks,colorlinks, urlcolor=blue,citecolor=blue,linkcolor=blue]{hyperref}
\usepackage{natbib}
\usepackage{verbatim}
\newcommand{\GG}[1]{}
\submitted{Accepted for publication in ApJ on July 17, 2018}

\shorttitle{Tidally induced morphology of M33}

\shortauthors{Semczuk et al.}

\begin{document}

\title{Tidally induced morphology of M33\\ in hydrodynamical simulations of its recent interaction with M31}

\author{Marcin Semczuk$^1$,
 Ewa L. {\L}okas$^1$,
  Jean-Baptiste Salomon$^{1,2}$,
  E. Athanassoula$^3$ and Elena D'Onghia$^{4,5,6}$}

\affil{$^1$ Nicolaus Copernicus Astronomical Center, Polish Academy of Sciences, Bartycka 18, 00-716 Warsaw, Poland \\
$^2$ Institut UTINAM, CNRS UMR6213, Univ. Bourgogne Franche-Comt\'e, OSU THETA,\\ Observatoire de Besan\c{c}on, BP 1615, 25010 Besan\c{c}on Cedex, France \\
$^3$ Aix Marseille Univ., CNRS, LAM, Laboratoire d'Astrophysique de Marseille, 13388 Marseille, France \\
$^4$ Department of Astronomy, University of Wisconsin, 2535 Sterling Hall, 475 N. Charter Street, Madison, WI 53076, USA \\
$^5$ Center for Computational Astrophysics, Flatiron Institute, 162 Fifth Avenue, New York, NY 10010, USA \\
$^6$ Vilas Associate Professor \\}

\begin{abstract}
We present a hydrodynamical model of M33 and its recent interaction with M31. This scenario was previously
proposed in the literature in order to explain the distorted gaseous and stellar disks of M33, as well as the
increased star formation rate in both objects around 2 Gyr ago. We used an orbit integration scheme to find which
estimate of the transverse velocity of M31 more favors the interaction scenario and then tried to reproduce it in our
simulations. M33 was modeled as a stellar and gaseous disk embedded in a live dark matter halo, while M31 was
approximated only with a live dark halo. In the simulations the two galaxies passed each other with a pericenter distance
of 37 kpc. Tides excited a two-armed spiral structure in the M33 disk, which is found to be the predominant spiral signal in the
observed galaxy and has been long known as a feature easily induced by tidal interactions. We found that the gaseous
warp produced by the interaction did not resemble enough the observed one and we performed an additional
simulation including the hot gas halo of M31 to show that this feature can be properly reproduced by
tidal forces and ram pressure stripping acting simultaneously on the gaseous disk. In addition to the spiral arms, tidal
forces produced the stellar stream similar to the observed one and triggered a star formation burst at
similar radii as it is observed.
\end{abstract}


\section{Introduction}

The Triangulum Galaxy (M33) is a late-type spiral and the third largest galaxy in the Local Group (LG) after the other
two more massive members of the LG, the Milky Way (MW) and the Andromeda Galaxy (M31). M31 and M33 form a
pair which is rather separated from the MW, with their relative distance $\sim 4$ times smaller than the
distance to the MW. Except for the spatial proximity, which results in proximity on the sky, there are several
observational hints pointing towards the possibility that the galaxies are gravitationally bound and have interacted
in the past.

\subsection{Observational evidence for the M33-M31 interaction}

The first morphological feature of M33 that might have been induced by the interaction is its gaseous warp. It was first
found by \cite{rogstad} and later several works (\citealt{corbelli97}; \citealt{putman}, hereafter P09;
\citealt{corbelli14}, hereafter C14; \citealt{kam17}) confirmed that the HI disk of M33 extends further than the stellar
component and is strongly warped in the outer parts. This warp results in the continuous twist of the position angle
(C14; \citealt{kam17}) and has a symmetric geometry, characteristic of tidally induced structures. However, the
geometry of the warp of M33 is a little peculiar, because it has the shape of a letter S, while the spiral arms, both in
the stellar and the gaseous disk, have a reversed chirality, i.e. a Z-like shape. Initially \cite{rogstad} proposed that the origin of the warp could be a primordial distortion of the
disk, that survived until today because the tilted rings would precess with different rates at different radii
(\citealt{kahn}; \citealt{hunter}). This scenario appeared more attractive than the interaction with M31, due to the
angular separation between the galaxies, which was believed to be too large to produce sufficient tidal forces.
However, a more recent study by P09 found that the gaseous features were very probably induced tidally by M31 in the
past 1-3 Gyr. They based this finding on the orbit analysis that was constrained by the measurements of the radial
velocities of M33 and M31 as well as the proper motion of M33 derived by \cite{m33pm}.

While the vertical distortion of the gaseous disk of M33 has been known since 1970s, the stellar disk was until very recently believed to be
unperturbed in this dimension. \cite{mcnature} reported finding a stellar structure extending
northwest and south from the disk of M33. This feature stretches up to three times further than the size of
the disk and has an orientation similar to the HI warp. \cite{mc10} and \cite{lewis} confirmed the alignment of the
orientations, however \cite{lewis} also pointed out an offset between the two structures. The lack of precise
overlay of both features was suggested to be due to the fact that the stellar component is only affected by tidal
forces while the gas could also have experienced shocking and ram pressure stripping (RPS).

Both the gaseous warp and the stellar stream-like distortion are features that strongly suggest some sort of past
tidal interaction. While they are not commonly found in spiral galaxies, M33 possesses another, more common
morphological trait that may have been induced by tides, namely the dominant grand-design spiral
structure. It is generally believed that M33 is a multi-armed spiral with a clear two-armed structure in the inner
stellar disk (e.g. \citealt{sm33a}; \citealt{sm33b}; \citealt{lia88}; \citealt{pitch93}).
It has been known for a long time that a grand-design spiral structure may be induced by tidal interactions. This
view is supported by plenty of observational evidence on galaxies like M51, where one of the two spiral arms is
pointing towards the flying-by companion (\citealt{m51like}). The scenario of triggering the two-armed spiral structure
was also confirmed in numerical simulations on several occasions. It was shown that spiral arms in a disky galaxy may
be induced by a passing-by smaller companion (e.g. \citealt{dobbs10}; \citealt{oh15}; \citealt{pettitt}; \citealt{elena16}) or during the
orbital motion around a bigger perturber, even of the size of a galaxy cluster (\citealt{bv};
\citealt{semczuk}). In these simulations the arms are triggered during the pericenter passages and later they
dissolve and wind up in about 1-2 Gyr.

Besides two-fold spiral arms, the stellar disk of M33 is also known to host a small bar (\citealt{corbelli07};
\citealt{mexicana}). While bars can be easily generated by global instabilities in cold  disks (for a review on bar
dynamics see e.g. \citealt{barRev}), it was shown that tidal interactions can also induce bar structures, e.g. in
smaller companions perturbed by more massive hosts (\citealt{l14}; \citealt{lokas16}; \citealt{gajda}).

In addition to the morphological features present in both the gaseous and the stellar disk of M33, a gaseous
bridge-like structure connecting M33 and M31 was observed by several authors. First, \cite{braun} reported finding an
HI stream that seems to join M31 with M33. The presence of this structure was later confirmed by \cite{lockman}.
\cite{wolfe} found that about 50\% of this structure is made of distinct clouds, while the rest is a diffuse component.
While some authors, like \cite{bekki} argued that the HI bridge may have originated from an interaction
between M33 and M31, \cite{wolfe} objected, stating that the collapse time-scale of these clouds ($\sim 400$ Myr) is much
shorter than the time since the hypothetical interaction ($\sim 1-3$ Gyr). More recently observations of \cite{wolfe16} cast even more doubt on the presence of the bridge, by
confirming that the majority of HI material in that region takes the form of discrete clouds.

The last argument supporting the interaction scenario arises from the analysis of the star formation histories (SFH) of
both galaxies. \cite{bernard12} obtained the SFH of M31 in the outer disk and reprocessed the fields studied by
\cite{barker} in M33. They found that a rapid increase in star formation rates (SFR) took place both in M33 and M31
about 2 Gyr ago. \cite{bernard12} concluded that these bursts could be triggered by a close passage of the galaxies and that a similar increase is consistent with state-of-the-art simulations of galaxy interactions and
mergers.

\subsection{Previous work}

Despite recent progress in high-resolution hydrodynamical and $N$-body simulations of galaxies, not many authors have
attempted to reproduce particular observed interacting systems. Among the most popular objects are M51 (e.g. \citealt{salo};
\citealt{theis}; \citealt{dobbs10}), the Antennae galaxies (e.g. \citealt{ant1}; \citealt{ant2}), the Cartwheel galaxy
(e.g. \citealt{cart1}; \citealt{cart2}) and the Magellanic Clouds (\citealt{besla12}; \citealt{elenafox}; \citealt{pardy}). Interestingly, M33 was not a popular target, in spite of (or perhaps due to)
its proximity and the wealth of observational data.

Inspired by the discovery of \cite{braun}, \cite{bekki} carried out simple test-particle simulations to verify whether
the discovered bridge-like structure could possibly originate from the interaction between M33 and M31. The outcome of
this numerical experiment was in favor of this scenario and he proposed that the interaction might have happened 4-8
Gyr ago. He also suggested that the HI warp of M33 might be fossil evidence of such a past interaction, however the
resolution of his simulation was not sufficient to study the detailed structure of the M33 disk.

The second attempt at modeling the M33-M31 interaction was motivated by the discovery of the extended stellar stream
by \cite{mcnature}. In the very same paper, \cite{mcnature} presented results of high-resolution $N$-body simulations
of M33 passing near M31. Both galaxies were modeled as exponential disks with bulges, embedded in dark matter halos.
The relative orbit had a pericenter of 53 kpc and tidal forces excited a warp in the M33 disk that wound up and
in projection closely resembled the observed distortion.

The simulations of \cite{bekki} and \cite{mcnature} were only constrained by the 3D velocity vector of M33, since the
radial and transverse velocities were known only for this galaxy. \cite{m33pm} obtained proper motions of M33 by water
maser observations. For M31 \cite{vdmG} estimated global transverse velocity by analyzing the line-of-sight kinematics
of its satellites. This estimate, however, was not used by \cite{mcnature} to constrain the mutual orbit. Later
\cite{sohn} used longtime {\it{Hubble Space Telescope}} (HST) observations of three fields in M31 to obtain its proper
motions. Their measurements were corrected by \cite{vdm12a} for the internal kinematics of M31. The obtained transverse
velocity of M31 was found to be $17\pm17$ km s$^{-1}$ and implied that M31 will merge with the MW in the future.
This measurement also had implications for the possible past orbit between M31 and M33. \cite{shaya} used Numerical
Action methods and integration backward in time to find the orbital history of the galaxies in the Local Group. They
found that the results consistent with the measurements of the proper motions of M33 and M31 suggest that M33 is now at
its closest approach to M31.

Recently \cite{Patel1} used backward orbit integration and showed that the assumption of the transverse velocity of
\cite{vdm12a} yields orbits with an unlikely recent ($\sim2$ Gyr ago) and close ($<100$ kpc) pericenter passage.
However, the robustness of the measurements by \cite{vdm12a} (and specifically the corrections for the internal motions
of M31) was recently questioned by \cite{s16}. Since astrophysical implications of the measurements of \cite{vdm12a}
were enormous, \cite{s16} separately estimated the velocity vector of M31, by modeling the galaxy and its satellites
as a system with cosmologically motivated velocity dispersion and density profiles. The resulting radial velocity was
consistent with the observed one, while the tangential component was surprisingly high, $\sim149$ km s$^{-1}$. This
conflicting result created the problem of which transverse velocity to adopt for M31 if one wanted to model the orbital
history of galaxies in the LG.

\subsection{This study}

In this paper we aim to show that observationally constrained structural parameters of M33 and M31
combined with a relative orbit that is consistent with measurements of the observed velocities (one or the other in the
case of M31) can reproduce the following traits of the interaction found in M33: the gaseous warp, the stellar stream,
2-armed spiral structure and an increase in SFH. In order to reach this goal we used high-resolution
$N$-body/hydrodynamical simulations. The model presented here does not attempt to reproduce exactly the history of the
interaction between both galaxies, rather, our aim was to show that combining observables with numerical methods can
result in structures similar to the observed ones, which in the past were often assigned to the interaction scenario.

The paper is organized as follows. In section 2 we present the orbit integration method that helped us to answer which
of the measurements of the transverse velocity of M31 favors more the interaction scenario. In section 3 we
give the details of the simulations (initial conditions, numerical methods and the adopted orbit) that were carried
out in order to reproduce the observed M33. Section 4 describes the properties of the simulated galaxy and its
similarities to the observed M33. Section 5 provides the discussion of our results and section 6 summarizes them.

\section{Orbit integration}

To construct a model of the interaction between M33 and M31 one requires mass models of both galaxies and their mutual
orbit. The orbit is constrained by the final relative position and velocity of the galaxies that can be derived from
observed sky coordinates, distances, line-of-sight (LOS) velocities and proper motions. Positions on the sky of both
galaxies are known very precisely. Measurements of distances and LOS velocities were carried out by different authors
(e.g. \citealt{gieren}; \citealt{vdmG} and references therein), and while errors can be smaller or bigger, they
converge to similar values. Proper motions of M33 were only measured once by \cite{m33pm}. Proper motions of M31 were
derived by several authors and some of the results are in conflict, which creates problems for the attempts to model the relative orbit of the M31-M33 system.

Two most recent and conflicting results for the proper motions of M31 were obtained by \cite{vdm12a} and \cite{s16}. In
order to verify which of the measurements better favors the scenario with a recent pericenter passage, we performed a
semi-analytic orbit integration similar to the one presented, e.g., in \cite{Patel1}. The adopted values of
relative 3D positions and velocities between the two galaxies are described in Appendix A. Computational details of the
orbit integration and parameters we used are briefly outlined in Appendix B.

Our approach to finding which one of the two transverse velocity estimates favors more the interaction scenario
consisted of three steps. In the first step we integrated the relative orbit of the M33-M31 system for 5 Gyr backward
in time, starting from the central values of the two sets of phase-space coordinates given by equations (A1) and (A2)
(hereafter called the vdM12 set) and then by equations (A1) and (A3) (hereafter called the S16 set). The orbit for the
vdM12 set had no pericenter in this time period and the current position was the closest approach. The orbit for the
S16 set had a pericenter passage about 2 Gyr ago, with a pericentric distance $>100$ kpc. Neither of these orbits was
satisfying, i.e. neither had the pericenter $<100$ kpc to presumably reproduce the observed morphology of M33, so we
performed additional orbit shooting.

In the second step we integrated multiple orbits forward in time, starting from modified values obtained in the first
step for the vdM12 and S16 sets. The modification was made by varying the magnitude and the direction of the initial
velocity vector. By changing these parameters (the magnification factor and the rotation angle) we obtained a grid of
possible orbits. From the calculated sample of orbits we selected those that had a pericenter passage closer than 100
kpc and checked how close they lie to the sets vdM12 and S16 after the pericenter passage. The proximity was defined in
terms of the $\chi^2$ statistics and lower values were found for the S16 set.
We used the orbit found in this way in the following simulations, however it turned
out to be slightly divergent from the one obtained by the orbit integration. The differences appeared after
including hydrodynamics and increasing the resolution of our simulations. The final orbit used in our fiducial model is
then a result of many iterative corrections and is discussed in greater detail in section 3.2.

One may argue that the method of selecting the orbit that favors the interaction scenario used in the second step is
incomplete since there is no guarantee that the range of initial magnitudes and directions of the velocity
vector can reproduce all the possible values of the observed positions and velocities allowed by observational errors.
Because of that, in the third step we tested our calculations with a simpler and more straightforward method and discussed
its results instead of the statistics of $\chi^2$ obtained in the second step. The method is similar to the one
presented in P09 and it consists of integrating orbits backward in time starting from values randomly selected from
the range allowed by the vdM12 and S16 sets. We randomly selected 15000 positions and velocity vectors for both vdM12 and
S16 from the range enclosed by the $1\sigma$ error bars and integrated orbits for 5 Gyr backward in time.

\begin{figure*}
\begin{center}
\begin{tabular}{@{}cc@{}}
\includegraphics[width=248pt]{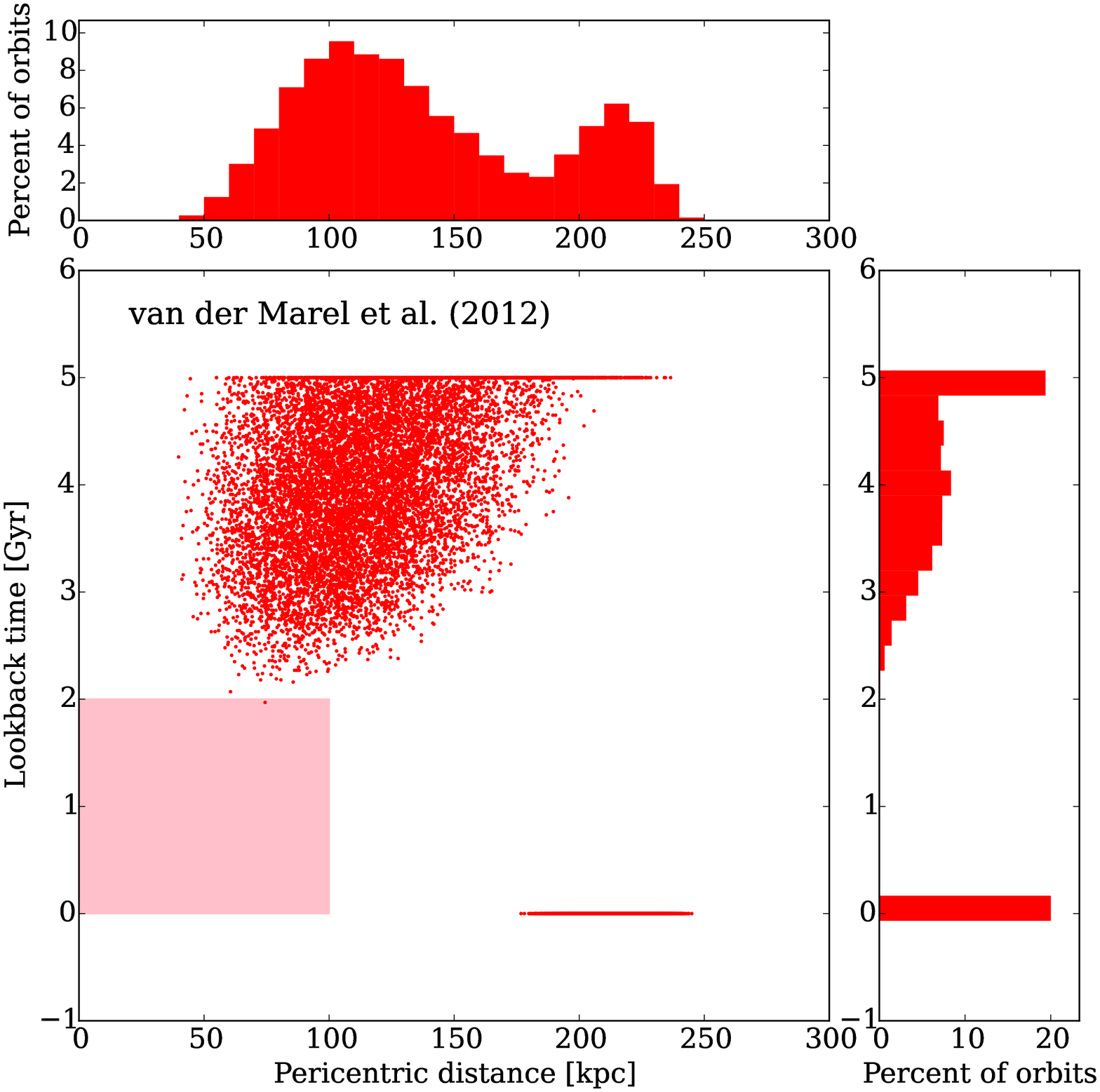} &
\includegraphics[width=248pt]{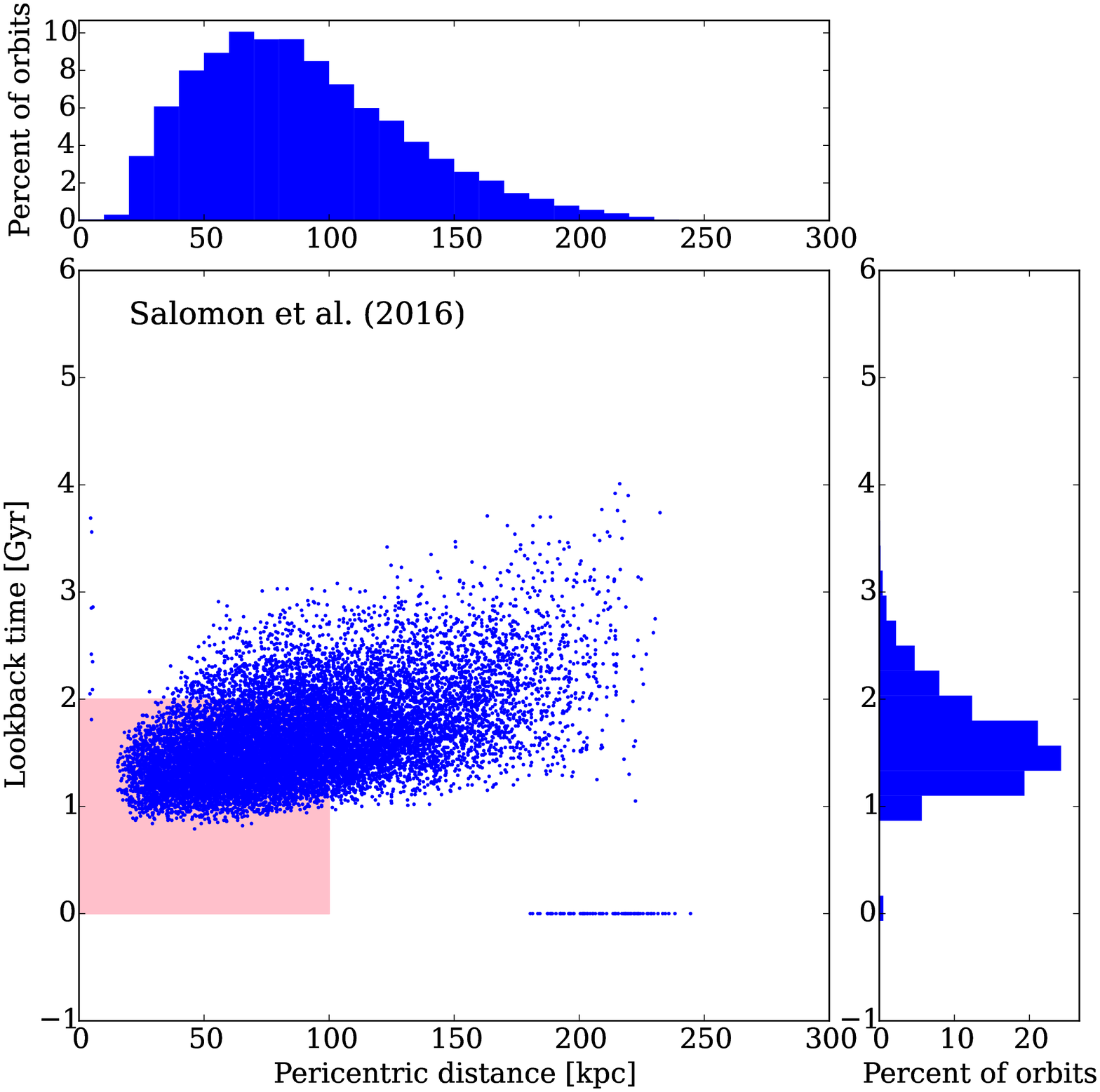}
\end{tabular}
\caption{
Distributions of pericenter distances and lookback times for orbits obtained via semi-analytic orbit
integration backward in time. The left panel shows the distribution for 15000 orbits that started from randomly selected
values from within error bars of relative position and velocity calculated by adopting the velocity of M31 derived by
\protect\cite{vdm12a} (equations (1) and (2)). The right panel shows the same results but after adopting values for M31 derived
by \protect\cite{s16} (equations (1) and (3)). The pink square in both plots indicates the area of values
($r_{\mathrm{peri}}<100\;\mathrm{kpc}$ and $t_{\mathrm{peri}}<2\;\mathrm{Gyr}$), suggested by
\protect\cite{putman} and \protect\cite{mcnature}, that would reproduce the observed morphological features of M33.}
\label{int}
\end{center}
\end{figure*}

Figure~\ref{int} shows the distributions of pericenter distances and lookback times at which the pericenters took place
for both initial sets of values. What we call the pericentric distances here are in fact just distances at the closest
approach. We did not integrate those orbits forward in time, or further backward, to really tell if it is a true
pericenter or if the pericenter is about to take place (or took place more than 5 Gyr ago). The analysis of Figure~\ref{int}
and especially of the horizontal histograms confirms our first findings that the S16 set favors more strongly the
interaction scenario than vdM12. Around 20\% of the orbits starting from vdM12 values are now at the closest approach,
while the majority of the rest had its closest approach more than 2 Gyr ago with only one point having a pericenter
recently ($<2$ Gyr ago) and closer than 100 kpc. Only around 0.5\% of the orbits starting from S16 are now at the
closest approach, while 82\% have a recent pericenter, i.e. less than 2 Gyr ago but not at the present time. 65\% of
all pericentric distances for S16 are smaller than 100 kpc. Therefore we conclude that measurements of proper motions
of M31 made by \cite{s16} better favor the interaction scenario than do the results of \cite{vdm12a}. In the rest of this
paper we will aim to reproduce in simulations the set S16 of the relative position and velocity between M33 and M31.

This finding is not very surprising, since the error estimates derived by \cite{s16} are $\sim$ twice as larger as those of
\cite{vdm12a} and naturally allow for more orbital configurations. Our findings are also partially in agreement with
the results of \cite{Patel1} and \cite{Patel2}, where one of the conclusions was that the interaction scenario is not
very plausible once one adopts the orbital values obtained by \cite{vdm12a}.

\section{The simulations}

\subsection{Initial conditions for the two individual galaxies}

\begin{figure}
\begin{center}
\includegraphics[width=240pt]{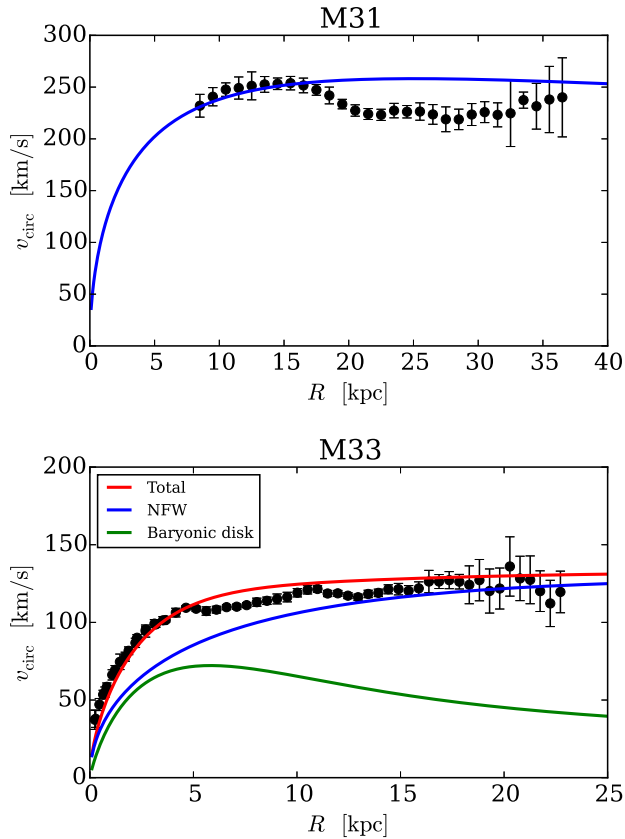}
\caption{Upper panel: the rotation curve of the initial model of M31. The observed rotation curve (\protect\citealt{m31mass}) is
represented by black dots. Lower panel: the rotation curve of the initial model of M33, after evolution in isolation
for 10 Gyr without star formation, feedback and cooling. The observed rotation curve (\protect\citealt{corbelli14}) is
represented by black dots.}
\label{rc_ini}
\end{center}
\end{figure}

Our aim when creating models of both galaxies is to reproduce their observationally derived
rotation curves (C14 and \citealt{m31mass}). This task is not straightforward since the interaction between the two
galaxies changes their structural parameters. This paper is devoted to the investigation of the changes that M33 might
have undergone due to such a hypothetical scenario and because of that we approximate M31 only as an NFW dark matter
halo with the virial mass $M_{\mathrm{M31}}=2\times10^{12}\;\mathrm{M_{\odot}}$ and concentration $c_{\mathrm{M31}}=28$.
These parameters were chosen to reproduce the rotation curve derived by \cite{m31mass} only with the dark matter
component (see the upper panel of Figure~\ref{rc_ini}). The baryonic content of M31 contributes more to the rotation curve
than the dark matter in the inner 10-15 kpc (see Figure 14 in \citealt{m31mass}) and therefore our adopted parameters
exceed the values of the parameters of the halo obtained by \cite{m31mass}, where the disk and the bulge were included
in the modeling of the mass distribution. Our model of M31 consisted of $2\times10^5$ particles.

Including the disk of M31 in the simulation would be an interesting possibility given the plethora of
substructure surrounding this galaxy (e.g. \citealt{lewis}). However, it would also make it more difficult to match the
inclinations of the two galaxies at the same time. The physical implication of not including the disk of M31 is the
spherical symmetry of its potential. The axisymmetric disk component would slightly bend the relative orbit between M33
and M31 out of the initial plane and this effect would have to be taken into account during iterative corrections
of the orbit.

Our model of M33 is more detailed and consists of three components: a dark matter halo and stellar and gaseous disks.
The initial dark matter halo had an NFW profile with  $M_{\mathrm{M33}}=5.2\times10^{11}\;\mathrm{M_{\odot}}$ and a
concentration $c_{\mathrm{M33}}=11$.
We increased the value of the halo mass with respect to what was estimated by C14, to take into account the
mass loss that will occur during the tidal stripping in the simulations.

The best estimate of the stellar mass of M33 provided by C14 is $4.8\times10^9\;\mathrm{M_{\odot}}$. The combined gas
mass of HI, $\mathrm{H_2}$ and helium is estimated to be $2.43\times10^9\;\mathrm{M_{\odot}}$ (C14). These values
summed up give a total baryonic mass of $7.23\times10^9\;\mathrm{M_{\odot}}$, with the stars contributing 66\% and the
gas 34\%. The surface density profiles of stars and baryons in total provided by C14 have a break at $\sim10$ kpc.
Because of this break, fitting one exponential stellar disk combined with the halo parameters and the gaseous disk will
not reproduce the given rotation curve to a satisfactory degree. One must divide the stellar disk into two components
with different slopes in order to have enough rotation both in the central and outer parts. In our simulations,
however, we decided to start with the simplest model and keep the number of parameters of the problem as
small as possible.
In order to preserve this simplicity and still reproduce
the rotation curve in the inner parts (and both inner and outer ones at the end of simulation, after tidal stripping of
the halo) we adopt the initial baryonic disk mass to be $M_{\mathrm{B}}=8.49\times10^9\;\mathrm{M_{\odot}}$ with a
scalelength $R_\mathrm{B}=2.5$ kpc and scaleheight $z_{\mathrm{B}}=R_{\mathrm{B}}/5=0.5$ kpc.

The initial conditions for the described disk, which consisted of $10^6$ particles, were generated using the procedures
provided by \cite{widrow1} and \cite{widrow2}. This method of generating initial conditions is suited for stellar disks
but not gaseous ones. We therefore divided this initial baryonic disk into stellar (35\%) and gaseous (65\%) particles
and evolved it in isolation so that the gaseous disk stabilized (similarly to \citealt{and2}). The simulation in
isolation was carried out with GADGET-2 $N$-body/smoothed particle hydrodynamics (SPH) code (\citealt{leapfrog};
\citealt{gadget2}). The galaxy was evolved for 10 Gyr with the initial temperature set to 5000 K. The sub-grid
processes were turned off, because the purpose of this run was to avoid gas instabilities and produce initial
conditions for the fiducial simulations, rather than study the evolution of the galaxy in isolation. During these 10
Gyr the gaseous disk thickened and stabilized. It became pressure supported instead of being supported by velocity
dispersion, as the stellar one is. The initial stellar and gas fractions (35\% and 65\%) that were kept constant in this
run were adjusted to reach the values inferred from C14 (66\% and 34\%) after the evolution with star formation
during the fiducial simulation.
This tuning was done via trial and error method by running simulations with subgrid physics and
iteratively checking which initial values lead to the final values close to the desired ones.

The initial surface density profile of the gas follows exponential distribution, with the same slope as the
stellar disk.  In general, the observed gas disk profiles are flatter in the center and do not follow an exponential
law, which is also the case for M33 (C14). This discrepancy is an artifact of the simple method of generating the
initial conditions that we used, nevertheless it has little influence on the model, since as shown in section 4.1 the
gas profile in the center will be significantly lowered and will mimic the observed one well, due to the conversion of
some of the gas into stars. 

The initial rotation curve plotted in the lower panel of Figure~\ref{rc_ini} seems to match the data in the
inner parts, while in the outer parts it overproduces the rotation. Note however that these values will be lowered down
as a result of tidal stripping of the dark matter halo. All the parameters of the initial conditions for both galaxies
are summarized in Table~\ref{tab1}.

\begin{table}
\begin{center}
\caption{Parameters of initial conditions for simulated galaxies}
\begin{tabular}{lll}
\hline
Components           & Properties          & Values                                 \\ \hline
M33 dark matter halo & Virial mass         & $5.2\times10^{11}\;\mathrm{M_{\odot}}$ \\
                     & Concentration       & $11$                                   \\
                     & Number of particles & $10^6$                                 \\
M33 baryonic disk    & Mass                & $8.49\times10^{9}\;\mathrm{M_{\odot}}$ \\
                     & Scalelength         & $2.5\;\mathrm{kpc}$                    \\
                     & Scaleheight         & $0.5\;\mathrm{kpc}$                    \\
M33 gaseous disk     & Number of particles & $6.5\times10^5$                        \\
M33 stellar disk     & Number of particles & $3.5\times10^5$                        \\
M31 dark matter halo & Virial mass         & $2.0\times10^{12}\;\mathrm{M_{\odot}}$ \\
                     & Concentration       & 28                                     \\
                     & Number of particles & $2\times10^5$                          \\ \hline
\end{tabular}
\label{tab1}
\end{center}
\end{table}

\subsection{The orbit}

The orbits derived by the orbit integration scheme described in section 2 happen to diverge from the orbits in the
$N$-body simulations with the same initial parameters and mass profiles. This divergence is due to the fact that the
orbit integration does not include tidal stripping and the resulting mass loss. In addition, the orbits found in
$N$-body simulations diverge from the orbits in SPH simulations, and those differ among each other for runs with
different particle numbers. All these differences are of the order of a few tens of kpc (up to $\sim 30$ kpc),
which makes it more difficult to accurately fit the adopted observed relative position and velocity (S16). It also
changes the pericentric distance (by $\sim 10$ kpc), which can affect the tidally induced morphological
features. Because of these numerical uncertainties, the orbit used in our fiducial simulation is not exactly the same
as the one derived in section~2, but instead it is a result of multiple iterative corrections that were made to best fit with the
morphological features (mostly the gaseous warp) and to end up as close as possible to the S16 set.

We applied a similar strategy to find the initial rotational angular momentum (spin) of our model of M33. The direction
of this vector is the same as the vector normal to the disk's plane, which can be derived from the observed inclination
and the position angle of M33. The first runs of our simulations were performed with the observationally derived
values; we found, however, that this vector was changing during the orbital evolution. Therefore we iteratively
corrected it, so that at the present time it reproduces reasonably well the observed position angle and inclination as
well as the observed morphology (we focused mostly on the gaseous warp).

The initial orientation of the
disk of M33 with respect to its orbit around M31 can be parametrized by two angles: $\alpha$, between the spin
of the disk and the orbital angular momentum, and $\beta$, between the spin and the direction of the velocity of
M33 on its orbit. We started with $\beta$ derived from the normal vector from observations and varied it by
$\pm45^{\circ}$ to see that both directions in the parameter space made the gaseous warp flatter. Hence we assumed
that for this parameter we are approximately in the best place. For $\alpha$ we tried eight different values from
the range between $\sim15^{\circ}$ and $\sim80^{\circ}$. Values close to the prograde case $\alpha=0^{\circ}$ keep the
tidal disturbance within the two dimensions of the plane of the disk and the warp-like distortions are not present.
Crossing $\alpha=90^{\circ}$ and approaching the retrograde case $\alpha=180^{\circ}$ decreases the effect of the tidal
perturbation of the disk and makes it hard to generate spiral arms (\citealt{resonantC}; \citealt{resonantB};
\citealt{resonant}). The final value of $\alpha$ turned out to lie only a few degrees away from what was derived from the
observations and we found that as long as it is not too close to $0^{\circ}$ it has very little influence on the 3D
appearance of the galaxy. We also found that rather than changing the inclination, other parameters were much more
influential in shaping the warp, namely the magnitude of the tidal perturbation (parametrized either by the
pericenter distance or by the mass profile of M31) and the inclusion of the hot gas halo of M31, which is discussed
later.

\begin{figure}
\begin{center}
\includegraphics[width=230pt]{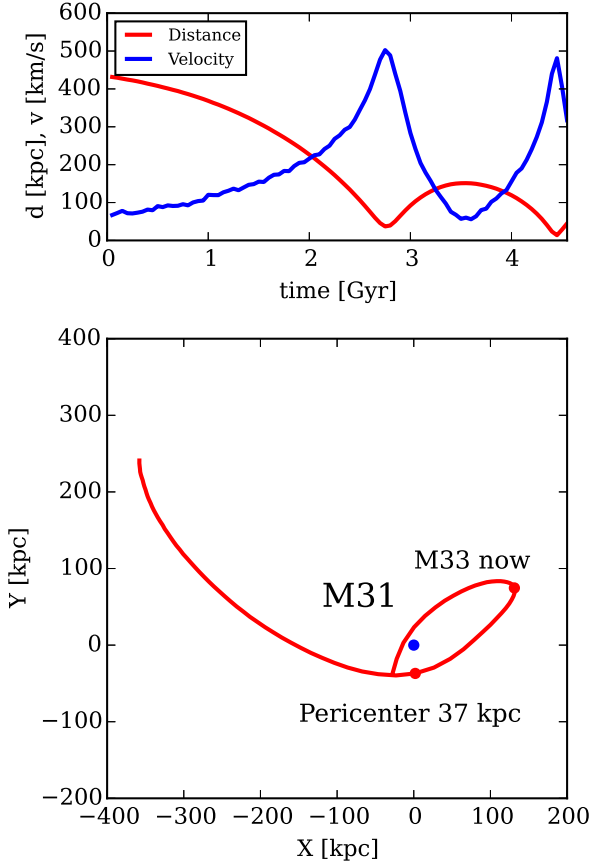}
\caption{Upper panel: the time dependence of the relative distance $d$ and velocity $v$ between M33 and M31 for the
adopted orbit. Lower panel: the shape of the relative orbit between M33 and M31 in the orbital plane. At each timestep
it was centered on M31.}
\label{orb}
\end{center}
\end{figure}

The values of the initial position, velocity and spin that we used in our fiducial simulation are summarized in
Table~\ref{tab2}. M31 was placed at the center of the coordinate system and its initial velocity was $(0,0,0)$ km
s$^{-1}$. The upper panel of Figure~\ref{orb} shows how the relative distance and velocity between M33 and M31 was
changing in time for the adopted orbit.  The orbit has a pericenter of 37 kpc at 2.7 Gyr after the beginning of the
simulation. The pericenter distance is larger than the sum of the sizes of the disks of both galaxies, hence the
approximation of the M31 potential as a pure dark matter halo does not imply that we neglect additional effects resulting
from the crossing of the disks. We note that dynamical friction decreased the apocenter of the orbit by a factor of
roughly 2.8, from 432 kpc to 151 kpc after the pericenter passage.

The lower panel of Figure~\ref{orb} shows the shape of the relative orbit in its plane. It is worth noting that
the position of M31 was changing significantly due to the attraction from M33 ($\sim100$ kpc), which is not so
surprising given that the adopted masses result in a mass ratio of M33 to M31 as large as $\sim26\%$. The orbit
presented here was centered on M31 at each timestep to make the image more clear.

\begin{table}
\begin{center}
\caption{Initial orbital parameters and inclination of M33}
\begin{tabular}{lllll}
\hline
Vector          & $X$    & $Y$   & $Z$    & Unit \\ \hline
Position        & -278.1 & 326.7 & 44.5   & kpc  \\
Velocity        & 56.5   & -6.5  & 32.8   & km s$^{-1}$ \\
Normalized spin & 0.048  & 0.998 & -0.038 & -    \\ \hline
\end{tabular}
\label{tab2}
\end{center}
\end{table}

\subsection{The code and the fiducial run}

The fiducial simulation presented in this paper was carried out with a modified version of the GADGET-2  $N$-body/SPH
code (\citealt{leapfrog}; \citealt{gadget2}) that includes star formation, feedback and cooling processes added as
described in \cite{hammer} and \cite{wang}. The sub-grid physics was implemented according to the recipes given by
\cite{cox}. We chose to adopt the values of the parameters as advised by \cite{cox}, namely a star formation
efficiency of 0.03, a feedback index of 2, a density threshold for star formation of 0.0171
$\mathrm{M_{\odot}}\mathrm{pc}^{-3}$ and a time-scale of feedback thermalization of 8.3 Myr. Instead of tuning the
sub-grid parameters listed above, we were changing the initial gas fraction to match the observed one at the present
time. If one intended to reproduce the simulations presented here with a code using a different sub-grid model, this
value would have to be tuned again to conform with the different sub-grid model, as it is generally believed that
sub-grid schemes can change the outcome of simulations very significantly (e.g. \citealt{jake}).

The implementation of the sub-grid physics using prescriptions of \cite{cox} includes cooling processes
described by procedures given by \cite{katz}. These procedures result in assigning every gas particle the neutral
hydrogen mass fraction. We stress here that throughout this paper we will use the name `gas' for gas particles in
general while the term `neutral hydrogen' or $\mathrm{HI}+\mathrm{H}_2$ will be
used for masses of gas particles reduced by this fraction given by the cooling procedures, in order to match
better the observations that mostly probe the HI content of M33.

We followed the evolution of the system for 4.5 Gyr with outputs saved every 0.05 Gyr. We adopted the softening
lengths of 0.09 kpc for the stellar and gaseous particles, 0.63 kpc for the dark matter of M33 and 10 kpc for the dark
matter of M31.

\subsection{The RPS experiment}
As described below, we managed to reproduce several features of M33 in the fiducial simulations. We found however
that one of these features, namely the gaseous warp cannot be reproduced to a satisfying degree of similarity
(especially at larger radii) by tidal effects alone. In order to argue that the RPS originating from the hot gas halo
of M31 would significantly improve the resemblance, we performed simulations with lower resolution that included the hot
gas halo of M31. The existence of the circumgalactic medium around M31 was demonstrated by \cite{m31hgh}, however there
are no strong observational constraints on the mass of the hot gas. Other galaxies are believed to have hot gas
halo masses roughly 100 times smaller than their dark matter masses (\citealt{mwhgh}). We started the modeling of the
hot gas halo with the mass value close to this estimate, however the final adopted mass was smaller than this, as a
result of tuning the effect of RPS to obtain the most similar gaseous warp.

In the simulation including the RPS, M31 was modeled as a Hernquist (\citealt{hernquist}) dark matter and hot gas
halo. Parameters of the dark matter component were fitted to reproduce the NFW profile discussed in section 3.1. and took
the values of $2\times10^{12}\;\mathrm{M_{\odot}}$ for the mass and 41 kpc for the scale radius. The hot gas halo had
the same scale radius and the mass of $3\times10^{9}\;\mathrm{M_{\odot}}$. The dark matter halo consisted of
$2\times10^5$ particles with the softening length of 10 kpc, and the gas halo was made of 70630 particles with
the softening of 0.2 kpc. Initial conditions for M31 for this simulation were generated using the CLUSTEP
code\footnote{https://github.com/ruggiero/clustep} (\citealt{clustep}). Initial conditions for M33 were generated
in the same manner as described in section 3.1 with very similar parameters, except for the particle numbers and
softening lengths. Particle numbers were $2\times10^5$ for dark matter, 126000 for the gas and 74000 for stars, while
softenings were 1.4 kpc, 0.2 kpc and 0.2 kpc, respectively.

\section{Properties of the simulated galaxy}

The initial model of the M33 galaxy was transformed during the simulated time by both sub-grid physics (star formation,
cooling and feedback) and tidal interactions with the M31-like dark matter halo. The most important morphological
features have been induced by the tidal interaction with M31 during the pericenter passage. To verify this we carried
out simulations with the same initial conditions for M33 but in isolation for 4.5 Gyr. A short discussion of this case
is included in the Appendix.

\subsection{General properties}
Star formation converts $2.61\times10^9\;\mathrm{M_{\odot}}$ of gas mass into stars
(hereafter called young stars, i.e. those stars that were born after the beginning of the simulation, unlike the old
stars that were present in the initial conditions), which resulted in a total stellar mass of
$M_{*}=5.59\times10^9\;\mathrm{M_{\odot}}$ (at the time of the best match; the selection of this time is explained in
section 4.2). This mass contributes $66\%$ of the baryonic mass, which is the ratio we were aiming to obtain,
consistent with the findings of C14.

\begin{figure}
\begin{center}
\includegraphics[width=230pt]{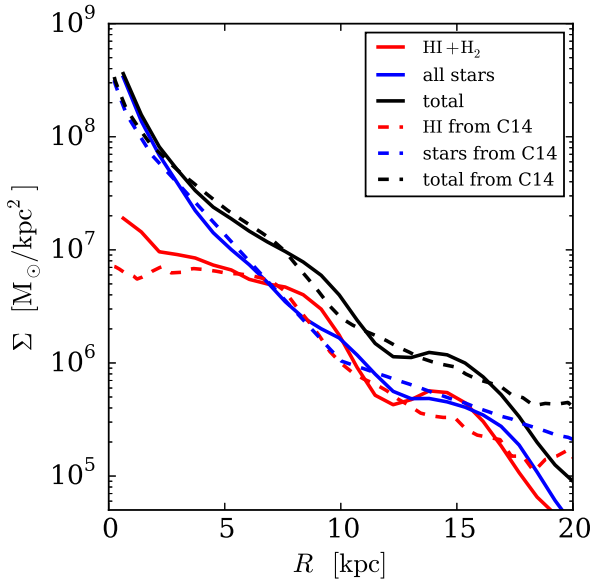}
\caption{
Surface density profiles for stars, neutral hydrogen and baryons in total in our model of M33
at the time of the best match, compared with observed density profiles for stars, HI and baryons in total from C14.}
\label{surf}
\end{center}
\end{figure}

The star formation created a central plateau in the surface density profile of the gas which is
also seen in observations (C14). It is clearly visible in Figure~\ref{surf}, where we compare the obtained
surface density profiles for stars, hydrogen and baryons in total with the observationally measured profiles given by
C14. We find that the simulated profiles are in a reasonably good agreement with the observed ones. We reproduced the
previously mentioned gaseous `core' and also approximated the slopes of the stellar and total baryonic distributions.

Tidal stripping does not strongly affect the baryonic mass of M33, however a significant fraction of the dark matter
mass was stripped. We fitted the NFW profile to the dark matter density distribution at the time of the best match and
obtained a mass of $M_{\mathrm{vir}}=4\times10^{11}\;\mathrm{M_{\odot}}$ and concentration $c=10.5$. These values are
within $1\sigma$ from the best-fit model of C14 and the mass is $\sim 77 \%$ of the initial halo mass.

\begin{figure}
\begin{center}
\includegraphics[width=240pt]{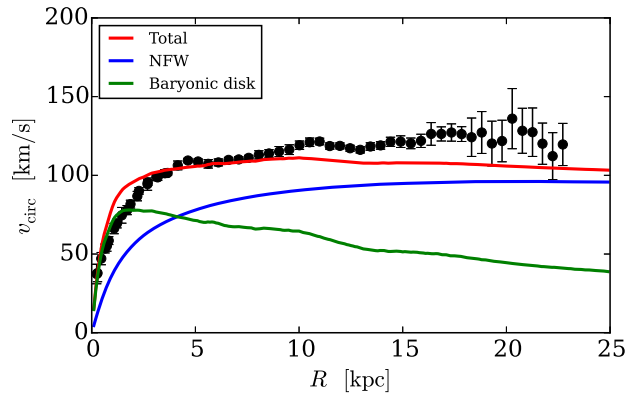}
\caption{The rotation curve of our model of M33 at the time of the best match compared with the observed rotation curve
provided by C14 (black dots).}
\label{rc}
\end{center}
\end{figure}

The rotation curve of M33 at the time of the best match is presented in Figure~\ref{rc}. This curve
was calculated directly from the sum of forces from particles
in simulations. We note that the initial curve presented in Fig~\ref{rc_ini} was calculated using analytical
formulae and was plotted only to show what we aim to reproduce. The obtained final rotation curve exceeds the observed
data by tens of km s$^{-1}$ in the inner $\sim 3\;\mathrm{kpc}$. This is due to the bar formation in this region, which
affects the potential of the galaxy. In the outer parts, the simulated curve falls slightly under the observed one,
which is a result of the tidal stripping of dark matter. However, despite these discrepancies on both ends, we find
that in general the simulated curve is a reasonable approximation of M33, being flat at around 100 km s$^{-1}$
throughout the majority of radii. The resemblance of the curve to the observed one is better when the bar is excluded
from the analysis and the disk component is calculated via the thin disk approximation from the fit to the surface
density profile that excludes the steep inner part originating from the bar.

\begin{figure*}
\begin{center}
\includegraphics[width=500pt]{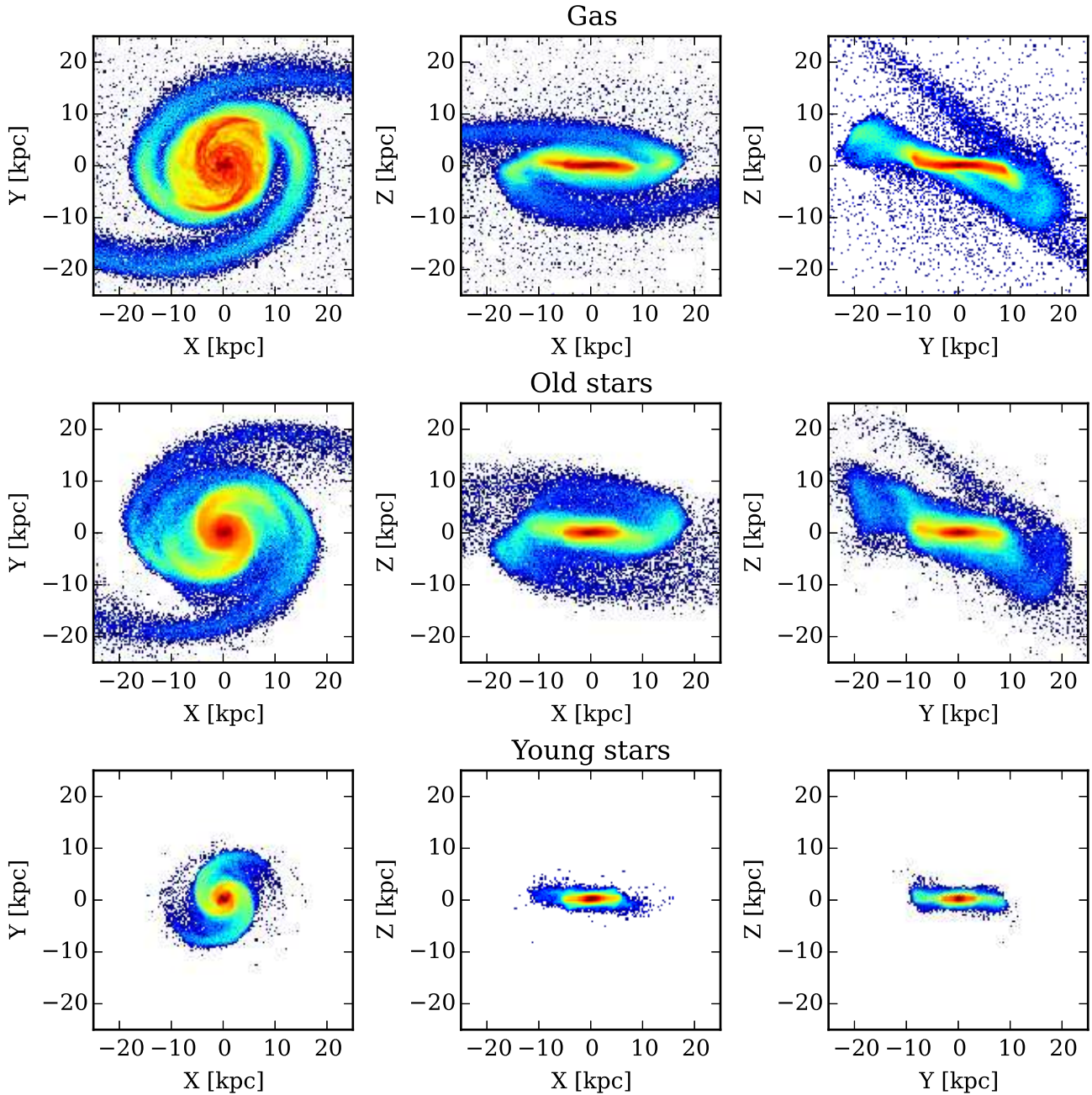}
\caption{Surface density distributions seen from three different directions for the gas, the old and young stellar
particles for the simulated M33 at the time of the best match.}
\label{map}
\end{center}
\end{figure*}

The tidal interaction during the pericenter passage induced spiral arms and a warp in both the gaseous and the stellar
disk. These features survived for 0.75 Gyr after the pericenter and are present at the time of the best match
(see Figure~\ref{map}). The two-armed, grand-design spiral structure characteristic of tidal encounters is present in
the gas, as well as the old and young stars. The off-planar tidal distortion (the warp) is mostly visible in the
gaseous disk and the old stellar disk. This is due to the fact that the new stars are born in the inner region of
the galaxy and the disk that they create is too small to exhibit a similar disturbance. The young stellar disk forms a small bar,
which is also present in the observed M33 (e.g.
\citealt{corbelli07}; \citealt{mexicana}). However, as discussed in section 4.5, the bar is not a tidal feature and
it also forms in isolation. Rather, the origin of the bar is related to the fact that the young stellar disk formed from
the gas by star formation is not stable against bar formation. As shown in \cite{lia13}, this process can be weakened
by increasing the gas fraction of the galaxy, however in our case the gas fraction is constrained by observations and
sub-grid physics.

\subsection{The warped gaseous disk}

Figure~\ref{map} clearly demonstrates that both the gaseous and the stellar disk have been warped by the tidal
interaction. \cite{mcnature} showed using $N$-body simulations that exciting such a distortion with similar parameters
of M33 and M31 is possible. Their model, however, was not constrained by the proper motions of M31 and did not
include gas physics. In this subsection we briefly describe how we found the best orbital epoch and the viewing
position that would result in the geometry of the warped disk similar to the observed one and we compare our simulated
images with those obtained by radio observations.

In order to find the projection and the orbital epoch that will result in a neutral hydrogen map resembling
best the observed one, we visually inspected images of the rotated gas disk of M33. The rotations were
constrained by the demand for the relative 3D position and velocity of M31 with respect to M33 to be consistent within
$2\sigma$ with the values of adopted (1) and (3) (reversed, because (1) and (3) were the relative coordinates of M33
with respect to M31 and here we look at the system from the perspective of M33). Using this procedure we found the
time of the best match to be 0.75 Gyr after the pericenter, which is 3.45 Gyr since the beginning of the simulation and
just 0.05 Gyr before the apocenter. The relative position and velocity of M31 in the reference frame corresponding to
the best match projection at this time are $\bm{X}_{\mathrm{rel,M31}}=(56.1,91.5,106.2)\;\mathrm{kpc}$ and
$\bm{V}_{\mathrm{rel, M31}}=(7.1,-38.8,41.1)\;\mathrm{km\ s}^{-1}$. One out of six of these coordinates is $1.75\sigma$
away from the adopted values. Another one lies within $1.25 \sigma$ and the remaining four are within $1 \sigma$.

\begin{figure}[t]
\begin{center}
\begin{tabular}{@{}c@{}}
\includegraphics[width=245pt]{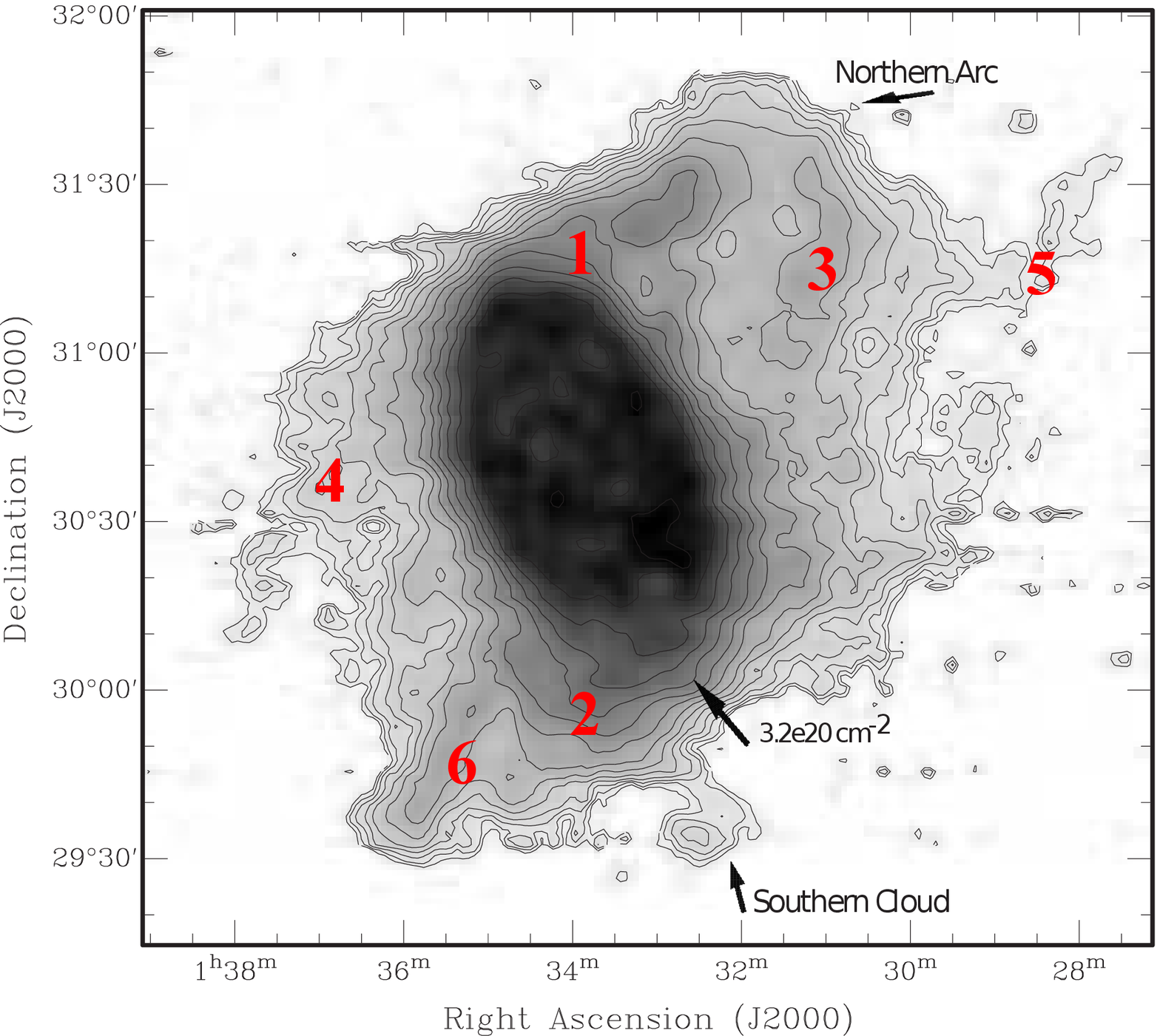} \\
\includegraphics[width=250pt]{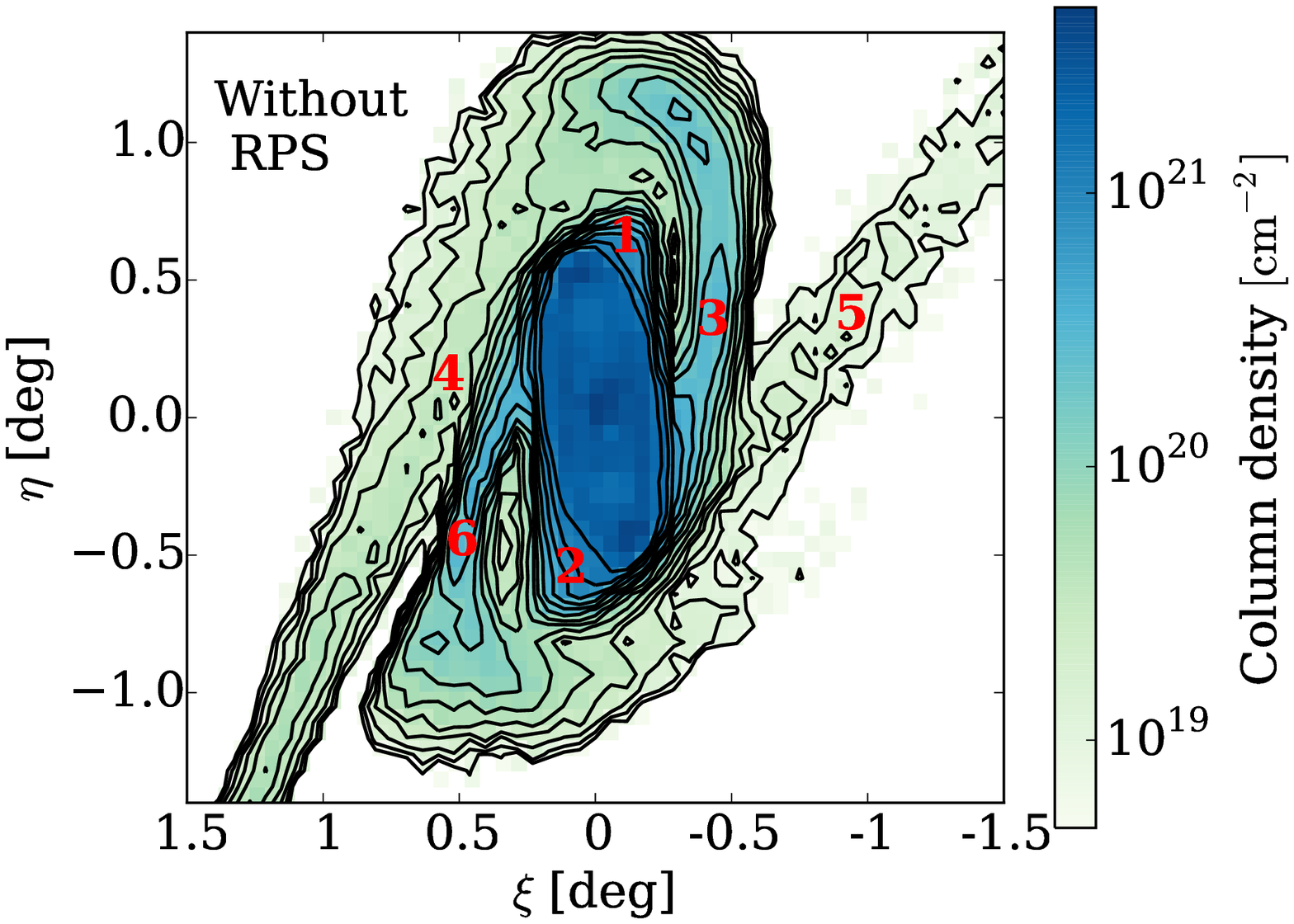} \\
\includegraphics[width=250pt]{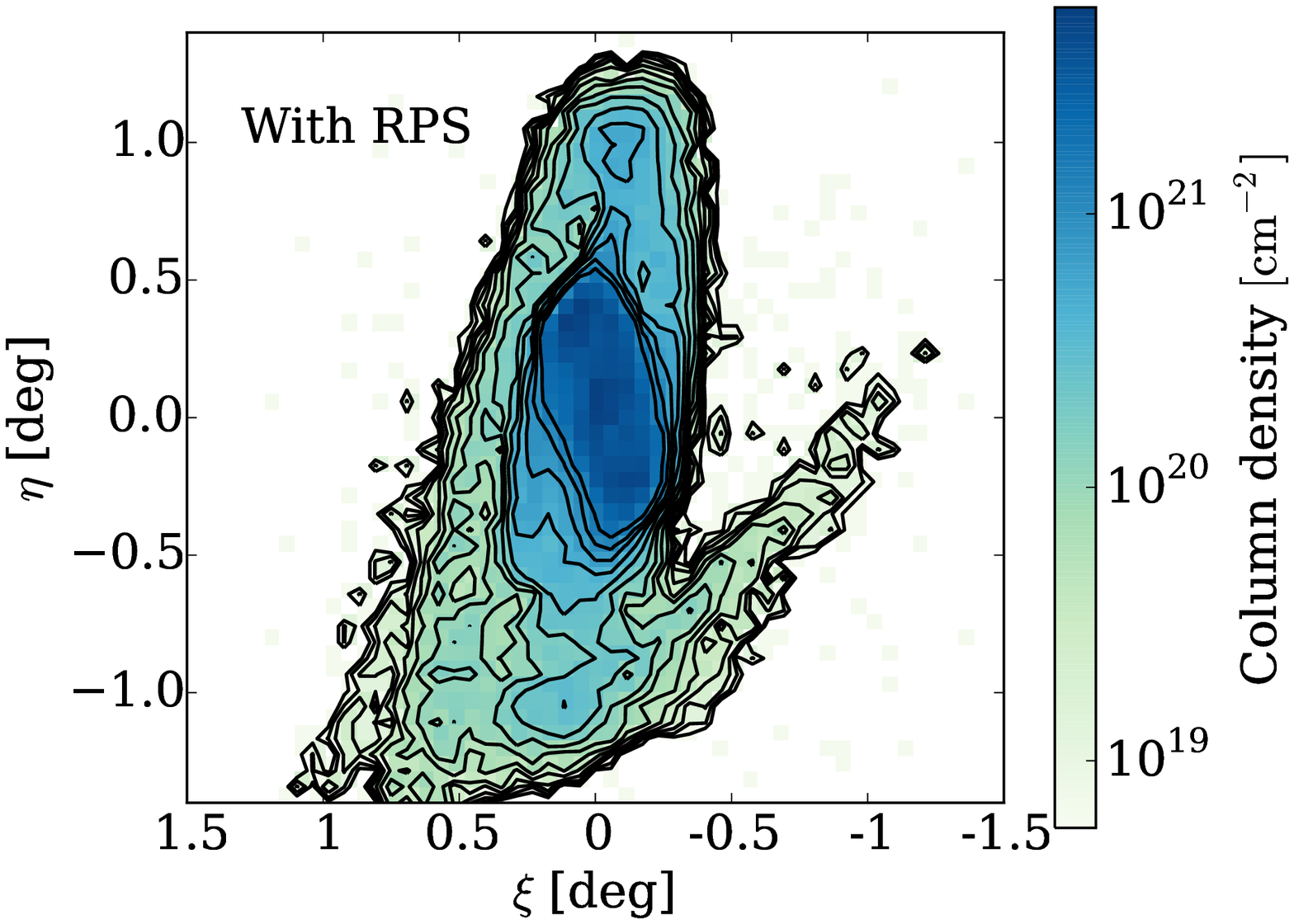}
\end{tabular}
\caption{Upper panel: the HI column density map of M33 published by P09 (\textcopyright AAS. Reproduced with permission). Middle panel: Best matching
neutral hydrogen column density map from our fiducial simulations. The isodensity contours were plotted at the same
levels as in P09. Red numbers mark tidally induced features that resemble corresponding features found in the image of
P09. Lower panel: Best matching neutral hydrogen column density map from simulations that included RPS from the hot gas
halo of M31. Inclusion of RPS seem to be necessary to weaken the strong tidal arms (3-6) that are almost invisible in
the observed galaxy.}
\label{warp}
\end{center}
\end{figure}

The neutral hydrogen density map for the best-matching time and projection from the fiducial simulation is presented in
the middle panel of Figure~\ref{warp} (coordinates ($\xi,\eta$) are distances in right ascension and
declination from the center of M33). The upper panel of the same Figure shows HI map published by P09. By comparing
both images we find that tidal interactions managed to only reproduce the inner warp of the gas disk, i.e. features
marked in red on both maps as 1 and 2. At larger radii in the simulated image, strong spiral/tidal extensions are the
most eye-catching feature, while in observations this is not the case. We identify several substructures in the image
of P09 that slightly resemble the spiral features in the simulated image (marked as 3-6), however in observations they
are merely composed of 1 or 2 differently shaped contours, while in the image from simulations they are the dominating
signal at these radii. Another discrepancy is that in the image from the fiducial simulation the gas structure is more
or less symmetric, while in the P09 image there seems to be more gas northwest from the galaxy. Also the northern
part of the warp marked as 1 is longer than its southern counterpart 2. This asymmetry in the observed gas is
pointing towards M31 on the sky and this coincidence motivated us to investigate whether RPS from the hot gas halo of
M31 may improve the simulated image.

The best-matching map from the simulations with RPS (found in the same way as for the fiducial simulation) is presented
in the lower panel of Figure~\ref{warp}. Strong spiral/tidal extensions present in the fiducial model (3 and 6) were
adequately weakened by RPS and the dominating signal is coming from the S-oriented warp. The resemblance to the
observations is still far from perfect (i.e. the galaxy is too elongated in simulations in comparison with more
flattened distribution in observations), but qualitatively the S-shaped structure is similar to the observed
one, with the inner spiral arms having the reversed Z-chirality. In the RPS image there is also too much gas at the
lowest density, located southeast from the disk. We suspect that this material (and maybe other discrepancies) may
arise due to the limitations of the SPH scheme used in the code to mimic the hydrodynamics of the gas. It is well known
that SPH codes fail the blob test (\citealt{agertz}; \citealt{gizmo}) and thus underestimate the RPS of the cold gas
and its mixing with the hot component.
Because of these numerical problems we decided not to explore the RPS
experiment in greater detail, since the physics of this process would not be properly modeled anyway, while the other M33 features can be reproduced simply with simulations of tidal evolution.
The asymmetry of the neutral hydrogen in the RPS image at first sight seems to be reversed
with respect to the observed asymmetry that motivated us to perform that experiment, however it is the case only for
the lowest density material. If we focus on the higher density, e.g. the 4th or 5th contour, where the gas physics is
better resolved, then the asymmetry is reproduced and the gas stretches up to $1^{\circ}$ north, while in the south it
only reaches $\sim0^{\circ}.6-0^{\circ}.7$.

\begin{figure}
\begin{center}
\includegraphics[width=240pt]{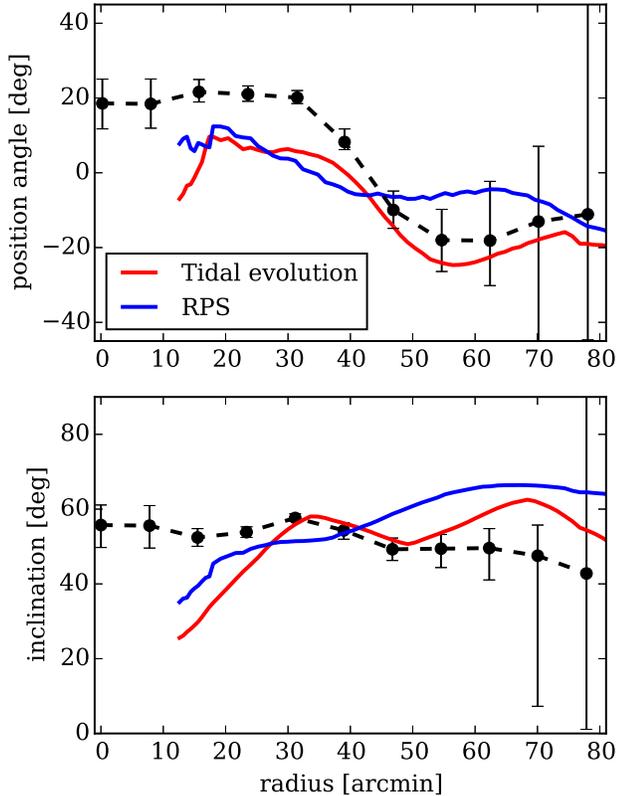}
\caption{Upper panel: the radial dependence of the position angle obtained for the best matching neutral
hydrogen density map from the fiducial simulation (red line) and the RPS experiment (blue line) in comparison with the
measurements from C14 (black points with error bars). Lower panel: the radial dependence of the inclination for the
same cases as in the upper panel.}
\label{pafit}
\end{center}
\end{figure}

\cite{corbelli97}, C14 and \cite{kam17} all reported that the gaseous warp quantitatively
manifests itself in the continuous change of the position angle of the disk with increasing radius.
The upper panel of Figure~\ref{pafit} presents the radial dependence
of the position angle obtained from images based on simulations and the comparison with the results from C14. The lines
from simulations were obtained by ellipse fitting to column density distributions. The curves from both simulations
exhibit a drop in values, the one from the fiducial simulations being more important, of the order of $30^{\circ}$, and resembling more the observed curve. The
drop from the RPS case is not so big, about $20^{\circ}$, and corresponds well with the impression given by the map in
Figure~\ref{warp}, i.e. that the overall shape of M33 is more elongated and the shift is milder than in observations.

Supplementary quantitative information about the warp is given by the radial dependence of the inclination
presented in the lower panel of Figure~\ref{pafit}. The curves from the simulations show
more structure than the one obtained by C14. This is due to the fact that the extended gaseous features are less
regular in simulations than in observations.
Contrary to the visual impression,
values from the simulations lie relatively close to the observed ones.
This coincidence is probably just the result of a very simple method of
deriving the inclination that we applied and a bigger discrepancy would be seen if we have fitted a tilted ring model as
in C14. Our models however are not very precise and do not require such sophisticated tools to analyze them; by
the ellipse fitting we just wanted to show that both the inclination and the position angle have approximately similar
values and radial dependence.

\begin{figure*}
\begin{center}
\begin{tabular}{@{}cc@{}}
\includegraphics[width=230pt]{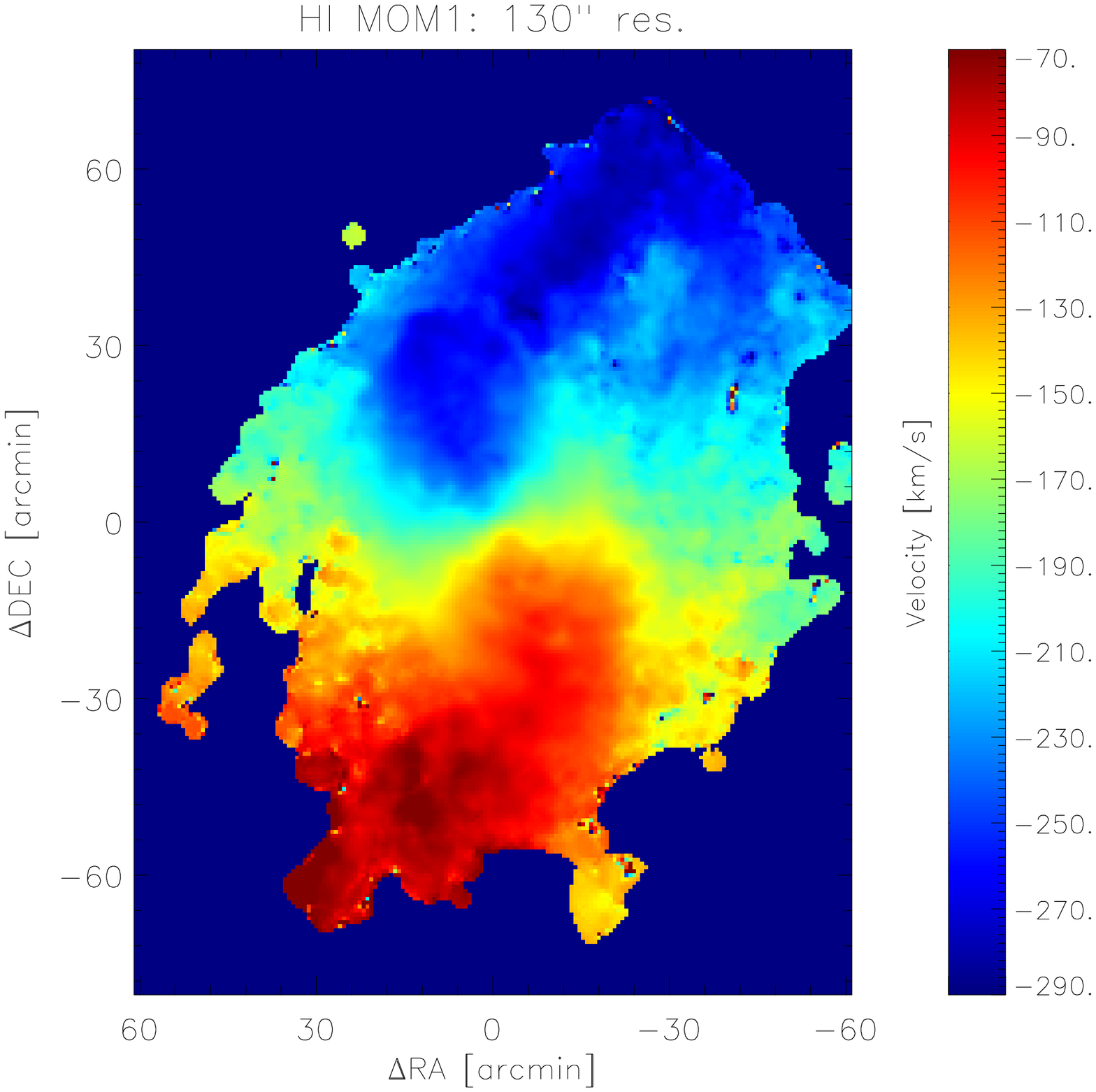} &
\includegraphics[width=240pt]{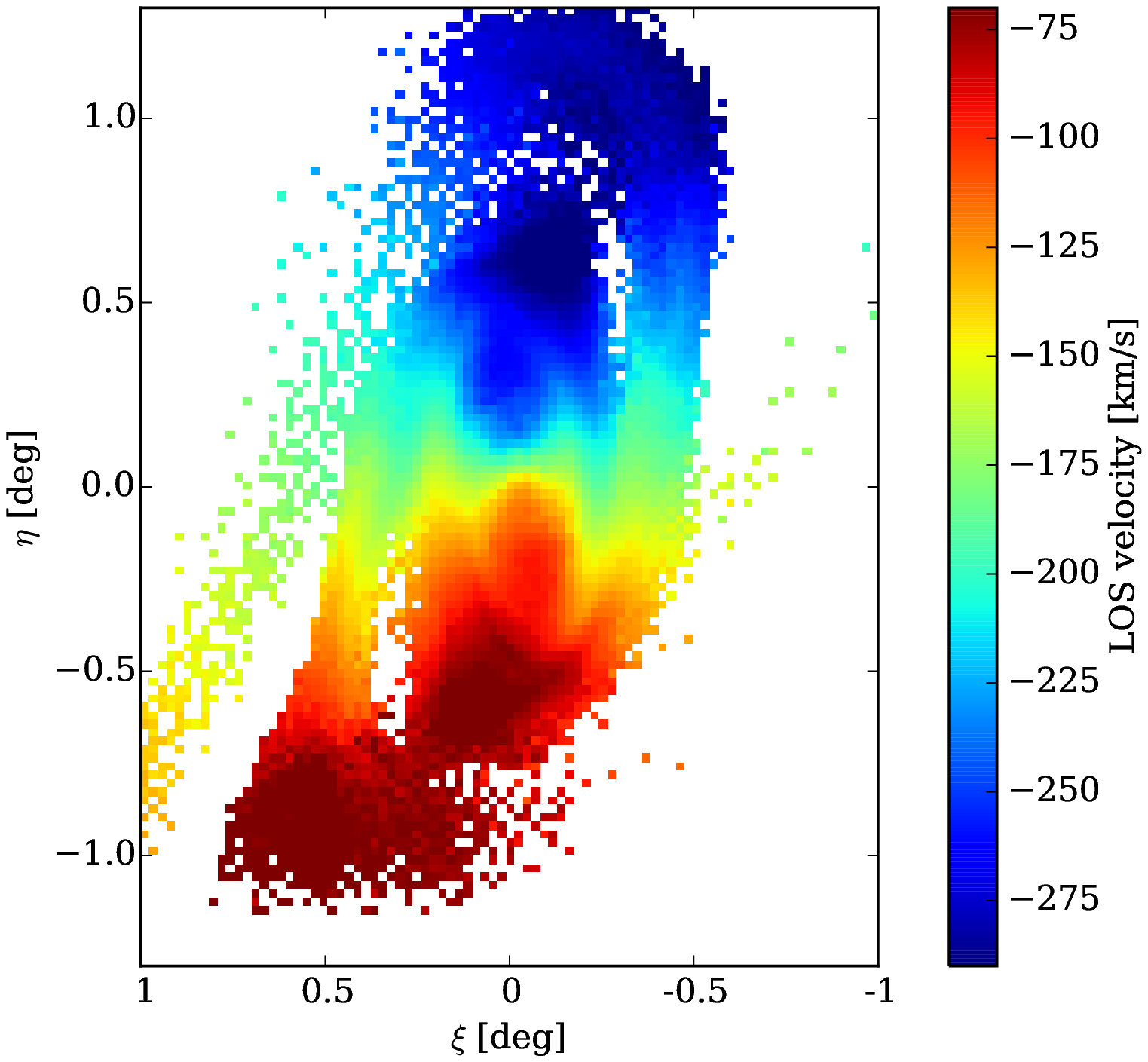}
\end{tabular}
\caption{Left: HI velocity map of M33 from C14 (reproduced with permission \textcopyright ESO). Right: Gas
velocity map of M33 for the best matching projection and time from our simulation. The simulated image has a
similar resolution as the one from C14. While constructing our map we took only those bins that contained more than
five particles in order to cut off the extended tidal tails and better mimic the image of C14.}
\label{kin}
\end{center}
\end{figure*}

The warp of the M33 gaseous disk manifests itself also in the kinematics of the gas. Figure~\ref{kin} presents the comparison between
the kinematic map published by C14 and the one obtained from our fiducial simulation at the time and the
projection of the best match.
The simulated image reproduces reasonably well the observed one and we find that the
distortion of the disk is clearly visible in the twist of the zero-velocity line, as was also found by other authors
(e.g. \citealt{kam17}). The image from fiducial simulations has a much thinner outline than the observed
one, which is the result of previously mentioned offset between the observed inclination and the one in the best match
projection. The best resemblance between the images is seen when one follows the lines that separate cyan from blue
and red from yellow. These clearly show where the disk ends and tidal features start to be visible. For example, the
tidal arm marked as 3 in Figure~\ref{warp} is seen in the simulated kinematic map as a red arc southwest of the main
disk. The corresponding arc is also present in the image from C14. The extended tidal arm is not clearly visible in the
HI map, however this red arc may be a hint that perhaps some sort of tidal arm is present
there. We note that the image obtained from the RPS experiment possesses features similar to those of
Figure~\ref{kin}.

\subsection{The stellar stream}

Unlike the gaseous warp, which has been known since the 1970s (\citealt{rogstad}), its stellar counterpart was discovered only
recently by \cite{mcnature}. Because this structure has been known for a much shorter time, it has been examined to a lesser
extent than the warp. In section 4.2 we selected the time and the best viewing position to match the warp. In
this subsection we briefly discuss the fact that for the same projection, the stellar stream is also present. In
general, we could have repeated the whole procedure to find a stream resembling the observed one more, however its
shape is not well-investigated as that of the warp and it changes its orientation from C-like to S-like when
selecting stars with different metallicities (\citealt{mc10}; \citealt{lewis}).

\begin{figure}
\begin{center}
\includegraphics[width=240pt]{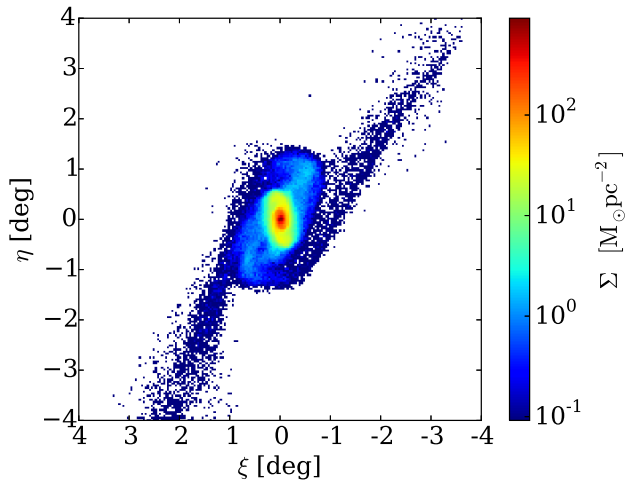}
\caption{The surface density distribution for all stars from simulated M33 at the time of the best match and seen from the
best viewing position.}
\label{stream}
\end{center}
\end{figure}

Figure~\ref{stream} presents the surface density distribution of all stars from our simulation. We decided to show the
simplest possible image that can be obtained from the simulations and not translate it into surface brightness or apply
complex processing to mimic images like Figure 13 of \cite{mc10}, since performing such procedures on the simulation data
would not be straightforward and require many assumptions and model-dependent methods. In this paper we only want
to quickly and qualitatively compare the resulting structure to observations.

We find that in general the simulated stream resembles the observed one. The two most evident features are in both
cases the northwestern and southeastern extensions. The simulated streams span $\sim2^{\circ}$, which is similar to the
observations. The structure discussed in \cite{mc10} seems a bit more vertically elongated, however this could also be
accomplished in the simulations by choosing a different viewing position.
The image from the simulation obviously has a better resolution and
some substructures that are visible here (for example a small gap in the southeastern extension) would probably be
smoothed out and not visible in observations. The surface density of the edge of the simulated disk is about
10 times bigger than the density of the stream. This agrees with the contours given by \cite{mc10}, where the ratio of
the surface brightness between the first and the last contour is 5.75. Apart from resembling the observed stellar
structure, our simulated stream is also very similar to the simulated one presented by \cite{mcnature}, where the
stream is made of wound up and projected tidal tails.

\subsection{The spiral arms}

\begin{figure*}
\begin{center}

\includegraphics[width=517pt]{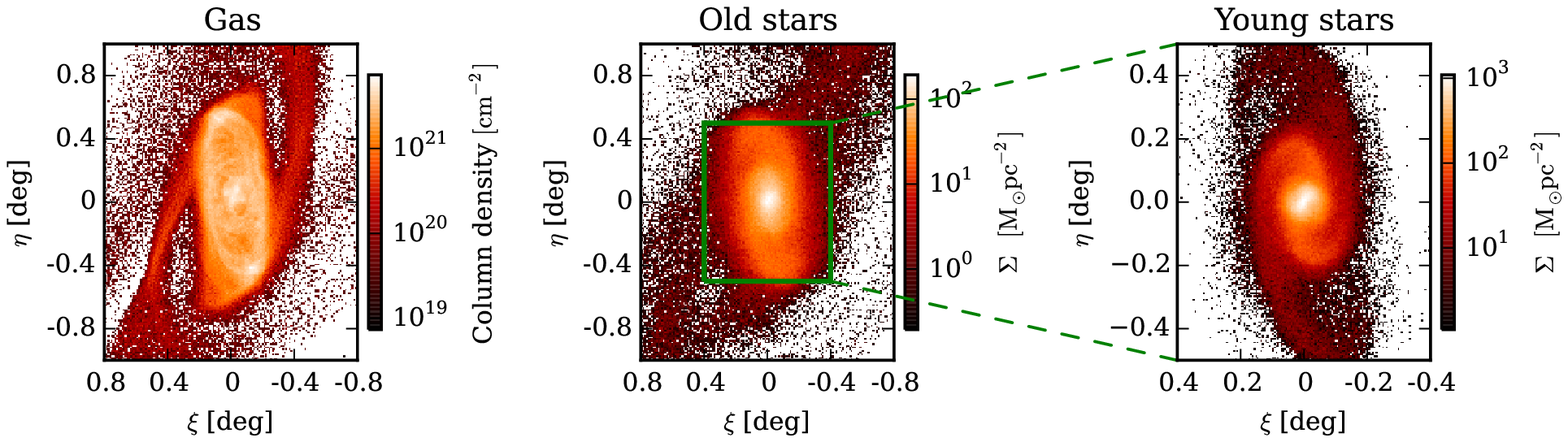}

\caption{Surface density maps of the neutral hydrogen (left), old stars (middle) and young stars (right) of the
simulated M33 at the best-match time and projection.}
\label{spirals}
\end{center}
\end{figure*}

As seen in Figure~\ref{map}, both the gaseous and the stellar disk possess grand-design spiral structure at the time of
the best match. The two-armed spiral signal is characteristic of tidally induced spirals and is also present in the
observed M33. Figure~\ref{spirals} shows how the spiral arms of our simulated galaxy appear in projection on the sky,
seen from the best-match point of view. For the neutral hydrogen map, the similarity with the HI image of C14 is
very good. The projection and the fragmented nature of a gaseous disk makes the arms look more flocculent, which is
also the case for the C14 image where it is hard to identify the number of arms.

In the stellar images the spiral arms are clearly more visible in young stars. This closely agrees with observations,
since generally young stars are a better tracer of spiral structure. The spiral arms reach deeper into the young
stellar disk than for the old disk, as a result of the fact that the young stellar disk was formed out of gas and is
about twice as small as the old disk (see Figure~\ref{map}). The spiral structure penetrates the young disk down to
$\sim3-6$ arcmin and is stopped there by the presence of a bar. The size of these spiral arms is similar to what is seen in
the B-V-I image of C14 made from the \cite{massey} survey. The connection between the arms and the bar is a bit
different in our simulated image compared to the one presented by C14. However, this results from the way the bar is
formed in our simulations (which is not by tidal interaction, see section 4.5) and it would be very difficult to match
its position angle with the observed one.

\begin{figure}
\begin{center}
\includegraphics[width=230pt]{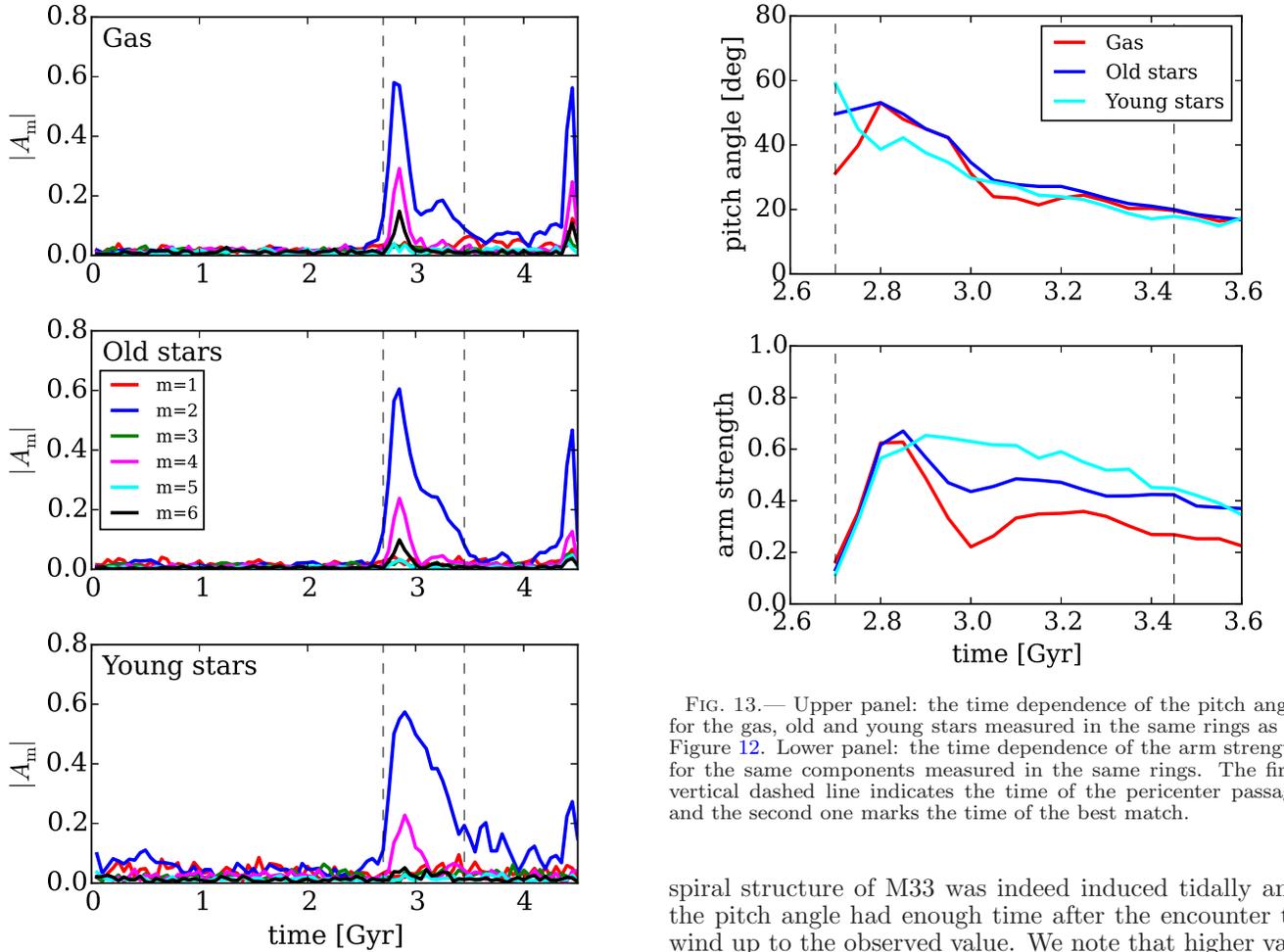}
\caption{Time dependence of $|A_m|$ Fourier coefficients measured in rings of $4\;\mathrm{kpc}<R<11\;\mathrm{kpc}$ for
the gas (upper panel) and old stars (middle panel) and $3\;\mathrm{kpc}<R<6\;\mathrm{kpc}$ for young stars (lower panel).
The first vertical dashed line indicates the time of the pericenter passage and the second one marks the time of the
best match.}
\label{am}
\end{center}
\end{figure}

Besides the qualitative inspection of the spiral structure, several quantitative properties may be measured for spiral
arms, namely their number (i.e. multiplicity), strength, pitch angle and pattern speed. First, we measured different Fourier modes $|A_m|$
(similarly to, e.g., \citealt{pettitt}) for the three different components to see what their dominant spiral signal is.
Their time evolution is presented in Figure~\ref{am}. A grand-design ($m=2$) signal is
clearly being tidally induced briefly after the pericenter passage and is stronger than other modes in all three
components. Unfortunately, we do not reproduce higher modes, which are present in the observed M33 (also in other
tracers, e.g. in H$\alpha$, see \citealt{kam15}). Reproducing the multi-armed structure in M33, which is not a dominant one,
would require better tuning of the initial model of M33, since the number of arms is mostly a function of a disk-to-halo
mass ratio (\citealt{lia87}; \citealt{elena2015}). In our work we focused mainly on tidally induced morphological features in M33
and we did not modify the model too much after we found the one that fits the rotation curve.

\begin{figure}
\begin{center}
\includegraphics[width=220pt]{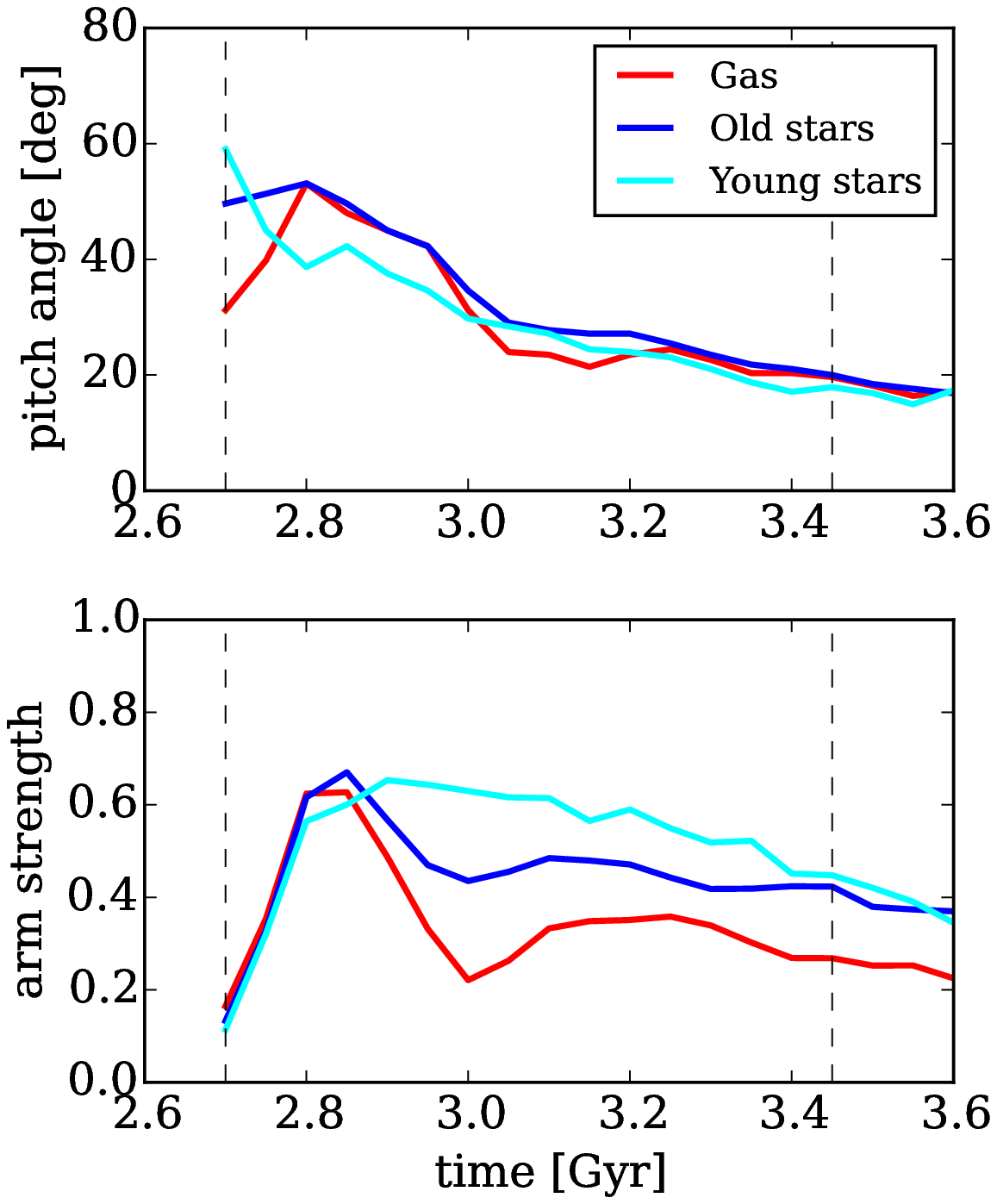}
\caption{Upper panel: the time dependence of the pitch angle for the gas, old and young stars measured in the
same rings as in Figure~\ref{am}. Lower panel: the time dependence of the arm strength for the same components measured in
the same rings. The first vertical dashed line indicates the time of the pericenter passage and the second one marks
the time of the best match.}
\label{ap}
\end{center}
\end{figure}

In order to measure the pitch angle and the strength of spiral arms we expanded the surface distribution of gas and
stellar particles in logarithmic spirals as discussed in e.g. \cite{sa86}. A more detailed description of the method we used can be found in section 3.2 of \cite{semczuk}.
The time evolution of these parameters after the pericenter passage is presented in Figure~\ref{ap}. The pitch angle
starts from high values during the pericenter, $\sim40^{\circ}-60^{\circ}$ for all components. Later it exponentially
decreases reaching $19^{\circ}.6$ for the gas, $21^{\circ}.8$ for old stars and $17^{\circ}.9$ for young stars, at the
time of the best match. The value for the best tracer of the spiral structure, i.e. young stars, agrees very well with
the pitch angle for the $m=2$ structure $16^{\circ}.5$ obtained by \cite{lia88} and $17^{\circ}.1$ found by
\cite{pitch93}. The values for other components also lie close to these measurements. This finding argues for the
possibility that the grand-design component of the spiral structure of M33 was indeed induced tidally and the pitch
angle had enough time after the encounter to wind up to the observed value. We note that higher values for the pitch
angle of the $m>2$ structure may be found in the literature (e.g. \citealt{sandage}) and also for $m=2$ measurements
may reach $\sim 40^{\circ}$ (references in \citealt{lia88}). However, as described by \cite{lia88}, these values
originate from measurements in the outer, looser spiral structure of M33 and the best fit for the brightest inner
two-armed structure is indeed $\sim16^{\circ}$. We conclude that the degree of winding up of the brightest component of
the spiral structure in our simulations agrees very well with the brightest component in observations.

The lower panel of Figure~\ref{ap} shows the time evolution of the arm strength.
It
peaks approximately at the same time as $|A_{m=2}|$ in Figure~\ref{am} and later it slowly decreases with time, which
means that the spirals are dissolving. At the time of the best match, the highest value of the arm strength is in the young
stellar component, which closely corresponds with the fact that the spiral structure is most pronounced in this component.

\begin{figure}
\begin{center}
\includegraphics[width=230pt]{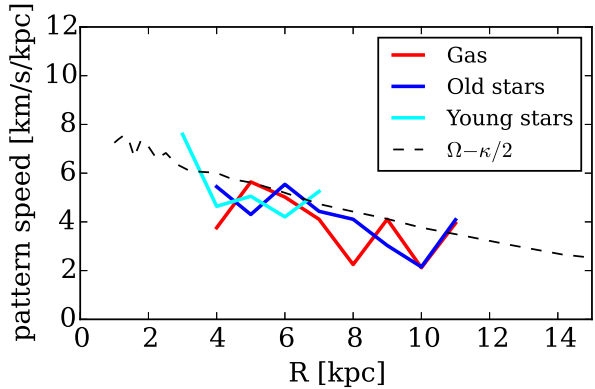}
\caption{The radial dependence of the pattern speed measured between the time of the best match and 0.15 Gyr before.}
\label{patt}
\end{center}
\end{figure}

The last property that we measured for the spiral arms of the simulated M33 is the pattern speed. We used the method
discussed, e.g., by \cite{dobbs11}, which relies on tracking the maximum of the surface density $\Sigma_{\mathrm{max}}$
in polar coordinates at a given radius between two time epochs.
The pattern speed can then be calculated as the difference between the angular positions of the maxima, divided by the time difference.
The radial dependence of the mean pattern speed of
the two-armed structure at the time of the best match is presented in Figure~\ref{patt}. We find that for both the gaseous
and the old stellar arms the pattern speed decreases radially and tightly follows the inner Lindblad resonance. This
indicates that the spiral arms are kinematic density waves, which is expected for tidally induced arms (\citealt{db14}).
Estimates of the pattern speed of M33 found in the literature (e.g. 15 km s$^{-1}$ kpc$^{-1}$ by \citealt{courtes71} or
28 km s$^{-1}$ kpc$^{-1}$ by \citealt{pitch93}) highly exceed the values obtained from our simulation. This discrepancy
probably arises from the fact that those estimates were based on the quasi-stationary density wave theory, where the
pattern speed of the arms is constant. This is not the case in our simulation and it was only recently reported
(\citealt{saha}) that constant pattern speed can be obtained at all in an $N$-body case.

\subsection{The bar}

\begin{figure}
\begin{center}
\includegraphics[width=245pt]{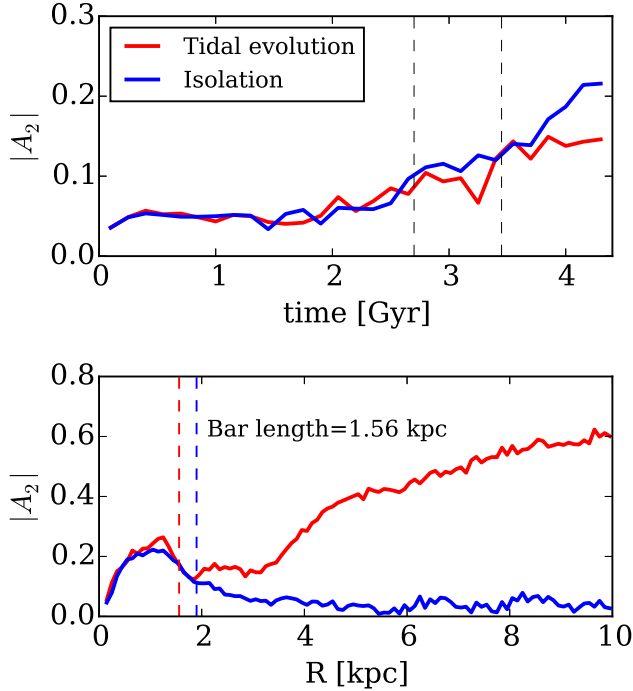}
\caption{Upper panel: the time evolution of the $|A_2|$ Fourier mode calculated within the radius of 3 kpc for our
fiducial simulation and for the case in isolation. The first vertical dashed line indicates the time of the pericenter
passage and the second one marks the time of the best match. Lower panel: the radial profiles of $|A_2|$ measured at
the time of the best match for the fiducial simulation and at the corresponding time for the isolated case. Dashed
lines indicate estimates of the bar length.}
\label{bar}
\end{center}
\end{figure}

The observed M33 is known to possess a small bar in the central parts of its disk (\citealt{bar1}; \citealt{bar2};
\citealt{corbelli07}; \citealt{mexicana}). As can be seen in Figure~\ref{map} and~\ref{spirals}, our simulated M33 also
forms a bar, mostly in the young stellar disk. In order to verify whether the origin of this bar is related to the
tidal interaction or due to the secular instability of the disk, we measured the time evolution of the Fourier mode
$|A_2|\equiv|A(2,p=0)|$ in the inner 3 kpc of the face-on surface distribution of all stars for the fiducial
simulation and the simulation of M33 in isolation. The upper panel of Figure~\ref{bar} shows that the tidal evolution has
very little influence on the bar formation and even slightly suppresses it.

The fact that the bar in isolation grows
bigger is also confirmed by the measurement of its length in both cases. We estimate the bar length by calculating the
radial dependence of $|A_2|$ and finding where it drops below a fixed threshold (here we assume it to be 0.5 of the
maximum value of the inner peak). This method yields a length of 1.56 kpc at the time of the best match for the
simulation with M31 and 1.9 kpc for the isolated case at the corresponding time (lower panel of Figure~\ref{bar}). While
this estimate (1.56 kpc) agrees quite well with one of the measurements in the literature (1.5 kpc by \citealt{bar1}),
it is at least two times larger than other measurements (e.g. 0.4 kpc by \citealt{bar2} or 0.7 kpc by
\citealt{corbelli07}). The value obtained by \cite{bar1} has been suspected of being biased by the influence of the very
tight spiral arms surrounding the bar in that region (\citealt{mexicana}) and the agreement with it seems to be
coincidental.

The bar in our simulation is formed in the young stellar disk, only because this disk is formed from
the gas and is very unstable. If we started our simulation later (or earlier) the bar length would be
different, since the bar would have enough time to grow bigger (as in isolation) or would not have enough time to grow
to the present size. In this paper we aim to investigate the very recent history of M33 and cannot determine when
exactly the processes responsible for the bar formation started. Our results only confirm that the influence of tidal
interactions on bars can be ambiguous, and they do not always accelerate bar formation (\citealt{bpettitt}).

\subsection{The star formation history}

\cite{bernard12} reprocessed the data from \cite{barker} for two fields in M33 at radii 9.1 and 11.6 kpc and found a
peak in the SFH that occurred $\sim2$ Gyr ago. The density of the SFR had a value of
$0.6\times10^{-9}\;\mathrm{M_{\odot}yr^{-1}pc^{-2}}$ at that time, which is approximately 3 times larger than the
average values found for earlier times. \cite{bernard12} also found similar activity in the SFH of M31 and suggested
that the past mutual interaction of both galaxies might have caused those synchronized peaks.

\begin{figure}
\begin{center}
\includegraphics[width=240pt]{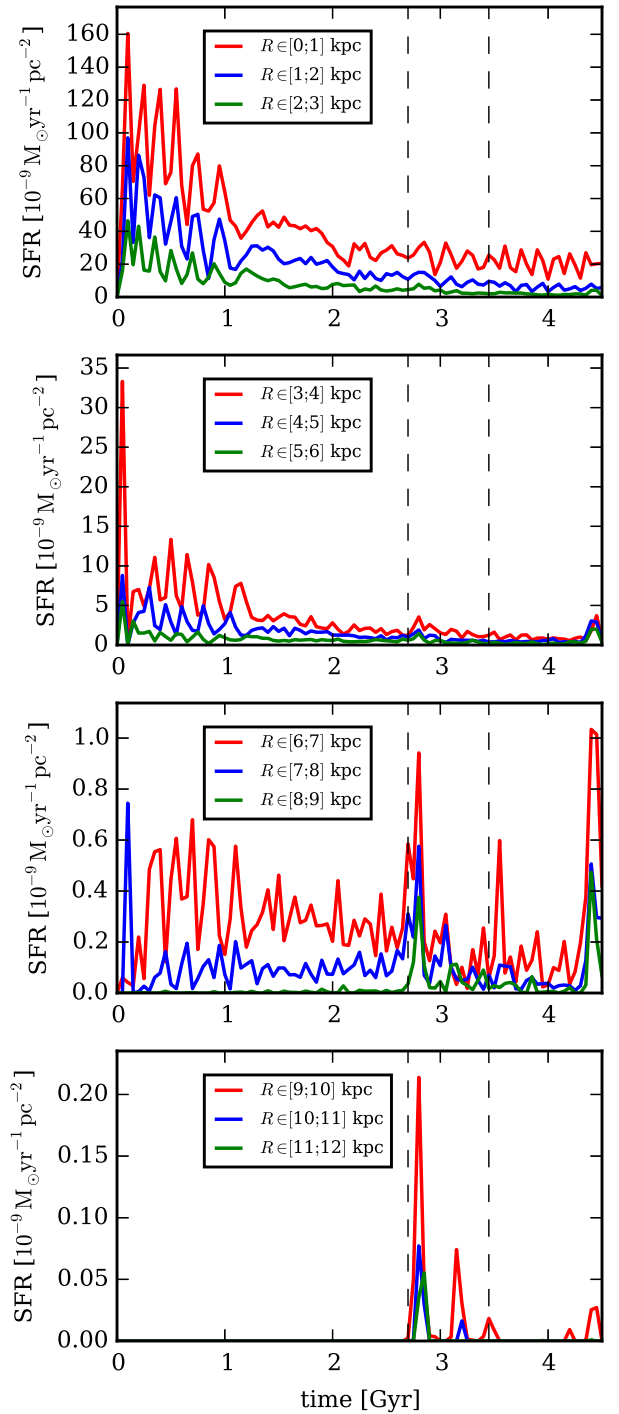}
\caption{The time evolution of the SFR normalized to the area of the ring (or circle) in which it was measured. The
first dashed vertical line indicates the time of the pericenter passage, the second marks the time of the best match.}
\label{sfh}
\end{center}
\end{figure}

We checked whether M33 in our simulation also exhibits a rapid increase in SFR at similar radii. Figure~\ref{sfh}
presents the time evolution of densities of SFR measured in rings (and one circle) of sizes of 1 kpc. At the beginning
of the simulation the values of SFR were obviously high in the central regions, where most of the star formation
happened. While the star formation slowly quenches, these values decrease and stabilize. At the radii from 6 to 12 kpc
SFR increases again rapidly after the pericenter passage (marked as the first vertical dashed line in Figure~\ref{sfh}).
The values are from 2 to 10 times greater than the average ones before. The values closest to those
found by \cite{bernard12} occur at distances from 6 to 9 kpc. However, it is difficult to
compare the absolute values between simulations and observations, because the SFR was measured in a different way and
in different kind of fields (here in rings, in observations in two square fields). The peak of the SFR in the simulation
is separated by around 0.75 Gyr from the time of the best match, a period $\sim 2.6$ times smaller than the one found by
\cite{bernard12}. When comparing these times of the peak it is worth keeping in mind that deriving the SFH is
strongly model-dependent. For example, for the same fields \cite{barker} obtained the increased SFR at $\sim3$ and
$\sim6$ Gyr ago. We conclude that the pericenter passage of $\sim 37$ kpc can induce an increase in the SFR of the same
relative order and at similar radii as observed.
We also found that introducing the hot gas halo of M31 does not change these results regarding SFHs at
different radii.

\subsection{The extended gaseous structure}

More than a decade ago \cite{braun} found a faint HI stream that seems to connect M31 and M33. Using test particle
simulations, \cite{bekki} showed that such a bridge-like structure may have been tidally induced by the interaction
between the two galaxies.
To see whether such feature would be present in the model of interaction presented here, we used the
simulations that included the RPS, since as we showed in section 4.2., this additional physical process is crucial in
reproducing better the outer regions of the cold gas of M33. We found no cold gas component stretching in a bridge-like
form between M33 and M31. This finding partially agrees with the results of \cite{wolfe16} who showed that the
majority of HI between the two galaxies is contained in small discrete clouds, rather than in a bridge. In our
simulations we do not have enough resolution to model such objects.

In our fiducial simulation we found cold gas stretching from M33 in the direction to M31 and also on the other side
of M33 (similarly to the stars as seen in Figure~\ref{stream}). This structure formed as a result of tidal stripping,
however, as discussed in section 4.2. we had to include RPS to weaken these features, since their signal was too
strong inside $1^{\circ}$ from the disk of M33, which is not seen in observations.

Despite not finding extended structure of neutral hydrogen north-west from M33, in the RPS experiment we find
low-density gas material southeast from M33 (partially visible in the lower panel of Figure~\ref{warp}). The
position of this material corresponds well with the HI clouds recently found by \cite{keenan}. The LOS velocities of
the material from the simulation and the clouds of \cite{keenan} are consistently oriented towards the observer,
however the observed material is approaching faster than M33 itself, while in our simulations this velocity is lower
than the one of M33. We did not investigate the properties of this material in greater detail, since one of the
conclusions of \cite{keenan} was that at the moment we cannot rule out the possibility that this gas is in fact some
contamination from the Magellanic stream.

\section{Discussion}

\subsection{Accuracy and uniqueness of the model}

As stated in the Introduction, the model presented here is not aimed to precisely reproduce the Triangulum Galaxy
and its present relative position and velocity with respect to Andromeda. The main goal of this study was
to show that once we adopt the values of structural parameters that are similar to the ones derived from
observations and use them in numerical simulations, the observed morphology of M33, the burst in its SFH and the
relative position and velocity with respect to M31 may be reproduced with reasonable accuracy. The purpose of this
paper was to show that the observed parameters allow for the interaction with M31 to sufficiently
disturb the disk of M33, and that such scenario utilizes an orbit that is consistent with measurements of distances and
velocities of both galaxies.

We noted several times that discrepancies between our model and the observations still persist. The biggest
discrepancy seems to be in the relative position between M31 and M33 at the best-match time and projection,
specifically in the $X$ direction, where it lies only within $1.75\sigma$ of the adopted error bar. This corresponds to
a difference of 41.1 kpc. The errors we adopted here as $1\sigma$ originated from the assumption that
measurements have Gaussian error distributions and we can easily transform them into different reference frames.
These errors are not the maximal possible errors. The maximal possible errors of the relative positions are the same in
all directions and equal to the sum of errors of both distances (assuming that the sky coordinates have exact values),
i.e. 63 kpc. This value is greater than the discrepancy of our results. However, the adopted errors that we used seem
slightly underestimated, when we compare them with a rather large range of observationally derived distances to both
galaxies, especially M33. For example, in the NASA/IPAC Extragalactic Database (NED) the measurements of the distance
to M33 range from $\sim620$ kpc to $\sim960$ kpc (with the mean value of 849 kpc and the standard deviation of 212 kpc).
This sample may be biased by very old measurements, nevertheless \cite{gieren} considered a sample of more recent
attempts and the range of distances was still quite extended, from $\sim720$ kpc to $\sim970$ kpc. Comparing this range
with our adopted error of 23 kpc suggests that taking such a small error may be recognized as a little too
optimistic. Reviewing the results presented in the literature and their discrepancies shows that the astrophysics of
the M33-M31 system is a branch of science with a distance accuracy of $\sim50-100$ kpc, rather than $\sim25$ kpc, as it is
for the projected position errors.

This huge range of possible values that are only weakly observationally constrained gives modelers a variety of
parameters to explore. The distance to M33 is not the only parameter with such a big allowed range. For example, the
mass of M31 may vary from $\sim7\times10^{11}\;\mathrm{M_{\odot}}$ (\citealt{m31low}) up to
$\sim2.5\times10^{12}\;\mathrm{M_{\odot}}$ (according to the introduction in \citealt{Patel2}). In general, one could
run simulations with different halo models of M33 and M31, gaseous and stellar disks of M33 and M31, inclusion or
not of bulges of both galaxies, the gaseous halo of M31, as well as different relative orbits and inclinations of both
galaxies, particle resolutions, hydrodynamical solvers and sub-grid models and their parameters. The simplest strategy
would be to perform simulations probing a multidimensional grid of parameters, however such a task would be
computationally expensive and the analysis of the simulations would be very difficult, given the abundance of
observational properties to reproduce.

While constructing the model described in this paper, we applied a different strategy, given the limited resources that we
had. In the first step we found the initial orbit using an orbit integration scheme. This orbit was later tested and
corrected in $N$-body runs. These runs were used as a basis for the following SPH simulations, which were also
iteratively corrected after comparing the obtained properties of the galaxy with observations. The model presented here
is a result of about 80 such iterations. Some of these intermediate steps reproduced certain observables better than the
fiducial model (e.g. the peak in the SFH occurred a longer time ago), however this was
obtained at a cost of less similarity to other observables. Reproducing a particular galaxy is therefore a
minimalization issue and the fiducial model we present here seemed to be a compromise in terms of the similarities
with the most important characteristics of M33.

It is certainly possible to obtain a better model. The simulation including the hot gas halo of M31
looks very promising and exploring this scenario further with a code that resolves the RPS better (\citealt{agertz};
\citealt{gizmo}) would be very interesting. Probing better the parameter space of orbits and inclinations in this setup
perhaps could also give a better result. Given the huge uncertainty of the proper motions of M31, we do not
exclude the possibility that the true velocity vector lies somewhere between the sets of vdM12 and S16.

Most of features of M33 that were induced due to the interaction with M31 in our simulation are of tidal origin
and therefore depend strongly on the orbit, inclinations and mass models. The sub-grid physics seems not to be very
important and our results are rather robust against different models of star formation etc. The peak in SFH after the
pericenter passage might be the only observable that could have been changed by sub-grid physics. We tested our
model with a code with different sub-grid prescriptions and a similar effect of the increase in SFR during the pericenter passage
was also found. Some extreme model with a very low density threshold and very high feedback could probably alter this
result and prevent strong increase in star forming activity.

\subsection{Comparison with previous works}

The first model of the interaction between M33 and M31 was presented by \cite{bekki}. The test-particle simulations
presented there were tailored to reproduce the HI bridge found earlier by \cite{braun}. In our model we do not find
such a bridge-like structure, as was the case in \cite{bekki}.
In the fiducial simulations without RPS we obtained instead tidal arms that were extended in both
directions up to $\sim4^{\circ}-5^{\circ}$ (similarly to stellar tails in Figure~\ref{stream}). These features however
were dissolved in the run with RPS that reproduced better the outer gaseous structures of M33. This discrepancy with
the results of \cite{bekki} comes from many differences between our works, however the most important is probably
the inclusion of hydrodynamics and hot gas halo in our case, since the run without it produced similar
features. Our finding of no bridge-like structure is in agreement with recent results (e.g. \citealt{wolfe}) that
question the possibility that HI between M31 and M33 is due to the past interaction between the two galaxies.
Apart from the bridge,
\cite{bekki} does not discuss other features associated with the interaction scenario, hence it is difficult to further
compare with his work.

The second model of the interaction was presented by \cite{mcnature}. They performed high-resolution $N$-body
simulations in order to reproduce the stellar distortion of the disk, the discovery of which was reported in the very
same paper. Similarly to \cite{bekki}, they adopted a small ($8\times10^{10}\;\mathrm{M_{\odot}}$) halo mass of M33
and a relatively large halo mass of M31 ($2.47\times10^{12}\;\mathrm{M_{\odot}}$). This assumption of a small mass
ratio of only $\sim3\%$ between M33 and M31 makes the modeling of the orbit much easier, since such a system can be
approximated as a small satellite orbiting a massive spiral. We assumed such masses of both galaxies with the aim to reproduce
their rotation curves (C14; \citealt{m31mass}) and this resulted in the mass ratio of $\sim 26\%$, for which the
system has to be treated more like two-body interaction of similar-sized objects.

Another difference between our work and \cite{mcnature} is that their orbit, just like Bekki's, was not constrained by
the measurement of the M31 transverse velocity, unknown at that time. Despite these different assumptions we find that
our model agrees with the one presented by \cite{mcnature}. The pericenter distance they used was similar ($\sim
50$ kpc) and the warp in their disk looks very similar to ours, especially for the stellar component
(Figure~\ref{stream}). The stream-like structure, just as in our case, is made of tidally distorted material that
wound up and in a specific projection has a different orientation than the disk. After the pericenter passage their M33
ends up more distant from M31 than ours (the apocenter is $\sim260$ kpc in their model and $\sim151$ kpc in
ours), but this is a result of the smaller mass ratio discussed earlier. Unfortunately, any quantitative results
regarding e.g. the spiral arms or SFH (due to the lack of hydrodynamics in their simulations) were not discussed by
\cite{mcnature} so we cannot compare our model with theirs in more detail.

Besides the work in the literature that aimed to reproduce the hypothetical past interaction between M33 and M31,
several authors have argued that such a scenario was unlikely using different orbital studies. First, \cite{shaya} used
the Numerical Action method and backward orbit integration to conclude that M33 and M31 are currently at the closest
approach. According to the information in the appendix of their paper, their scheme did not include the effect of
dynamical friction. However, as shown in Appendix A of \cite{vdm12b}, orbital studies with and without
implementation of the dynamical friction may vary a lot (i.e. apocenter may change from 800 kpc to 100 kpc), with the
former agreeing much better with the simulations.
After finding that M33 is now at its closest approach to M31, \cite{shaya}
suggested that Andromeda XXII (And XXII), the only candidate for a satellite of M33 (\citealt{and22}), could
have disturbed the disk of M33. We will discuss this alternative scenario more in section 5.3.

More recently, \cite{Patel1} and \cite{Patel2} explored the past orbital history of two systems, the MW-Large Magellanic
Cloud and M31-M33. In the first paper (\citealt{Patel1}) they applied backward orbit integration to find possible
orbits of both systems and compared them to orbits from dark matter only cosmological simulations. In section 7.1 of
their paper they state that from backward orbit integration of 10000 initial velocities drawn from $4\sigma$ proper-motion error space they found that less than 1\% of orbits had a pericenter smaller than 100 kpc in the last 3 Gyr. As
can be seen in Figure 13 of their paper, these 10000 velocities cover well the area surrounding the velocity that results
from adopting the \cite{vdm12a} proper motions, while the area near the results of \cite{s16} is not probed.

In section 2 of our paper we extend these results of \cite{Patel1} to also probe the region surrounding estimates of
the velocity resulting from adopting the values of \cite{s16}. Our findings agree with the conclusion of \cite{Patel1}
that once one adopts the transverse velocity of M31 from \cite{vdm12a} the recent and close pericenter passage is not
very plausible. This agreement also holds if we take into account the fact that we used different mass models of the
galaxies (in particular M33 is almost twice as massive in our work). The conclusions about the recent pericenter
change once the estimates of \cite{vdm12a} are replaced by those of \cite{s16}. Since the publication of the most
recent estimates of the M31 transverse velocity, no other measurements were made to support one or
the other of the conflicting estimates. Furthermore, cosmological simulations (\citealt{carlesi}) do not strongly exclude
any of these estimates. Moreover, such simulations do not exclude a close pericenter passage, as
discussed by \cite{Patel1}, where the analysis of orbits from the Illustris-Dark (\citealt{illustris}) simulation showed
that orbits with pericenters even lower than 55 kpc are very common ($\sim30\%$) for similar pairs of halos.
The probability of orbits with pericenters smaller than 50 kpc was found to be even higher in a recent
study of orbits of similar pairs in cosmological simulations that included baryonic physics (\citealt{cautun}).

In the second paper \cite{Patel2} used a Bayesian inference scheme and the Illustris-Dark simulations to estimate
the masses of MW and M31.
One of the conclusions of this paper was that after adopting the criteria for the recent interaction scenario, the mass
of M31 would be $\sim10^{12}\;\mathrm{M_{\odot}}$. This value is not far from the estimates of the M31 mass found in the
literature, for example it lies within the error bars of the result of
\cite{watkins}, $1.4\pm0.4\times10^{12}\;\mathrm{M_{\odot}}$. We argue that the precision of the estimates of the mass
of M31 is not high enough to rule out $\sim10^{12}\;\mathrm{M_{\odot}}$ and abandon the interaction scenario.

\subsection{Alternative scenarios}

Both \cite{shaya} and \cite{Patel1} proposed a passage of And XXII as a new alternative scenario that could disturb the
gaseous and stellar disk of M33.
Here we try to roughly estimate the impact of the tidal force that such a passage would have had and compare it with the
impact in the scenario presented in this paper. \cite{elmegreen} defined the parameter $S$ that quantifies the strength
of the tidal interaction. Their definition makes use of the masses of the perturber and the perturbed galaxy, the
pericentric distance, the size of the perturbed galaxy as well as the timescales of the interaction and of motions of
the stars in the perturbed disk. Since constraining the orbit of And XXII around M33 would give too many free
parameters, we decided to apply instead a parameter that does not use the timescales and was discussed e.g. in
\cite{oh15}:
\begin{equation}
	P=\bigg(\frac{M_{\mathrm{ptb}}}{M_{\mathrm{g}}}\bigg)\bigg(\frac{R_\mathrm{g}}{d} \bigg)^3.
\end{equation}
Here $M_{\mathrm{ptb}}$ and $M_{\mathrm{g}}$ denote the masses of the perturber and the perturbed galaxy,
$R_\mathrm{g}$ is the size of the perturbed galaxy, which we take to be 5 times the disk scalelength, and $d$ is the
pericentric distance. For the case of the simulation described in this paper, the tidal parameter takes the value of
$P=0.03$, where we assumed that the mass of the perturber is the mass of M31 enclosed within the pericentric distance.
We know from our simulation that the tidal interaction characterized by this number can disturb the disk of M33 and the
induced morphology mimics the observed one to a certain degree.

We now want to find out what size of pericenter And XXII would need to have in order to provide the same tidal impact.
In order to estimate this we need the mass of And XXII. Adopting the value of \cite{shaya},
$1.3\times10^7\;\mathrm{M_{\odot}}$, we obtain the pericenter of 1.3 kpc. \cite{eric} estimated the mass in And XXII
enclosed within the half-light radius to be $\log( M_{1/2}/\mathrm{M_{\odot})}=6.67\pm1.08$. If we take the upper limit
of this measurement and assume that the half-light radius encompasses only 10\% of the total mass, the estimated
pericenter will still be only 4.5 kpc. Such small pericenter values suggest that the low mass And XXII would have to
pass extremely close to M33 in order to have the tidal impact similar to the one induced by M31 in our simulation. We
expect that such a close passage would perturb more the inner parts of the disk of M33, which seems to be undisturbed.
The warp and the stellar stream occur at much greater radii. We find that the scenario in which And XXII induced
the warp in M33 is less likely, due to the small mass of this satellite.

Another possibility is that more massive satellites of M33 disturbed its outer disk and they are
yet to be discovered. This scenario could perhaps explain the warped gaseous feature, since gas is less massive and
hence less bound and easier to distort. A different possible scenario for the gas disk distortion could also be an
asymmetric gas accretion. However, these two explanations would not be very convincing in the case of the stellar
component. Studying MW analogs, \cite{gomez} found that a satellite should have a mass of at least 1\% of the host to
induce some sort of vertical distortion in the disk. It would be very surprising if a satellite of M33 of a
corresponding brightness existed and was not yet discovered. Perhaps the distorted stellar and gaseous disks are
remnants of a major merger that M33 underwent in the past, but this scenario would raise another question of why the
inner disk remained unaffected by such an event. Hopefully, future observations and simulations will shed more light
on the peculiar history of M33.

Very recently, \cite{gaia} utilized Gaia DR2 results to estimate proper motions of both M31 and M33. Based on
the agreement of the new result for M31 with their previous HST estimate and on the orbit integration sample from
\cite{Patel1} they argue that M33 may be on its first infall into M31. We find this interpretation unlikely for the following three reasons. Firstly, the error bars of \cite{gaia} estimates are very large and the
resulting 3D relative velocity between M31 and M33, based solely on Gaia DR2 measurements, is in $1.75 \sigma$
agreement with the relative velocity for the best match of our model (section 4.2), in which the two galaxies recently
interacted. Secondly, the orbit integration sample from \cite{Patel1} was selected from $4\sigma$ space around \cite{vdm12a}
measurements and naturally will fall far from the new measurements (in Figure 4 of \citealt{gaia}), as that region was not fully probed and therefore the
distance between the new results and points with a pericenter might be just an artifact of the probing of the phase
space. Finally and most importantly, we do not consider a semi-analytic orbit integration method to be a robust
computational tool that can rule out with any confidence one or another scenario for the past history of M33 and M31.
Orbits obtained by this method do not reproduce the simulations accurately, since it does not include mass loss
originating from tidal stripping and parameters of the dynamical friction have to be fitted to a particular setup.

Moreover, in the close proximity to M31 there are some massive satellite galaxies that would most likely have
some influence on the M33-M31 orbit. The mass of M32 was estimated to be $8\times10^{10}\;\mathrm{M_{\odot}}$
(\citealt{m32}), which is around 20\% of the mass of M33, and M110 has a similar magnitude to M32 (\citealt{m110}).
Including these factors in the orbit integration would create a lot of degeneracy, since their proper motions are
unknown and finding the true orbit between M33 and M31 would be even more complicated. The simple scenario of the interaction that we discussed here favored more the \cite{s16} estimates, but perhaps the influence of M32 and M110
would change this conclusion and other estimates of transverse velocity of M31 would be in agreement with the M33-M31
interaction scenario.

\section{Summary}

In this work we revisited the scenario discussed in P09 and \cite{mcnature}, proposed to explain the disturbed
stellar and gaseous disks of M33 by a recent passage close to M31. We used the orbit integration method to verify which
of the measurements of the transverse velocity of M31 favors more this scenario. We found that while the estimates
of \cite{vdm12a} do not support this common history of the two galaxies, the estimates of \cite{s16} allow for it. We
performed $N$-body/SPH simulations aiming to reproduce the observed disturbed morphology of M33 and at the same time to be
consistent with the 3D relative position and velocity of the galaxies resulting from the estimates of \cite{s16}. The
fiducial setup presented here fulfills the orbital conditions and had the pericenter at the distance of 37 kpc which was
close enough to tidally disturb the disk of M33.
Mass models of both galaxies in our simulations were constructed to roughly reproduce the observed
rotation curves.

We found that the tidal impulse originating from such interaction is sufficient to excite two-armed spiral
structure similar to the one found to be the dominant spiral component in the observed M33. Tides also induced
distortion in the stellar and gaseous disks at larger radii, with the former having the shape and the extent
similar to the observed one. The disturbance of the gaseous disk, however, was found to be similar in the inner
parts while the outer were dominated by strong tidal/spiral features that are not present in the observed data. We
showed by performing an additional run including hot gas halo of M31 that the RPS is a crucial component in modeling
the gaseous warp of M33 in greater detail.

Finally, we also found that the tidal forces in our simulations were sufficient to compress the gas in M33 during the
pericenter passage and trigger a burst of star formation at the similar radii as found by \cite{bernard12} and
hypothesized to be due to the passage near M31, since similar activity was found in its SFH at approximately the same
time.
The model presented here did not aim to reproduce the observed M33 in a great detail, and this was not
achieved, but rather to demonstrate that observationally constrained structural and orbital parameters of the system
allow for the interaction to trigger in M33 features similar to the observed ones.

\section*{Acknowledgements}

This work was supported in part by the Polish National Science center under grant 2013/10/A/ST9/00023. We are grateful to an
anonymous referee for useful comments that helped to improve the paper. We thank L.
Widrow for providing procedures to generate $N$-body models of galaxies for initial conditions. We are grateful to S.
Fouquet for sharing the modified version of the GADGET-2 code and his collaborators F. Hammer and J. Wang for
incorporating the modifications. We thank S. Z. Kam and L. Chemin for sharing the HI data of M33 as well as E. Corbelli,
M. Putman and the editorial board of A\&A for granting us the permission to reprint certain figures. Acknowledgments
are also due to J.-C. Lambert for his tutorial about the tool for the visualization of the simulations glnemo2 and for
making it public. We appreciate insightful discussions with G. Gajda, K. Kowalczyk, I. Ebrova, N. Peschken and J. S. Gallagher III that contributed to this paper. MS is grateful for
the hospitality of Laboratoire d'Astrophysique de Marseille and the University of Wisconsin at the time of his visits. 
EA acknowledges financial assistance from the CNES and access to the HPC resources of
CINES under the allocation 2017-[A0040407665] made by GENCI.
ED acknowledges the support and the hospitality of the Center for Computational Astrophysics (CCA) at the Flatiron Institute during the preparation of this work and the Vilas Associate Professor Fellowship at the University of Wisconsin, Madison.
This research has made use of the
NASA/IPAC Extragalactic Database (NED) which is operated by the Jet Propulsion Laboratory, California Institute of
Technology, under contract with the National Aeronautics and Space Administration.

\appendix
\section{A. Relative positions and velocities}

Following \cite{vdm12a}, we adopted the distance to M33 and its error as $D_{\mathrm{M33}}=794\pm23\;\mathrm{kpc}$
(\citealt{m33dist}) and for M31 $D_{\mathrm{M31}}=770\pm40\;\mathrm{kpc}$ (\citealt{vdmG} and references therein). We
combined these values with the distance of the Sun from the Galactic center $R_0=8.29\pm0.16\;\mathrm{kpc}$
(\citealt{solar0}) to obtain the relative position vector between M33 and M31 in the so-called Galactocentric rest
frame. This Cartesian reference frame $(X,Y,Z)$ is defined as follows: the center lies on the Galactic center,
the $X$-axis points in the direction from the Sun to the Galactic center, the $Y$-axis points in the direction
of the Sun's rotation in the Galaxy and the $Z$-axis points towards the Galactic north pole. The resulting relative
position vector between M33 and M31 with error bars is
\begin{equation}
\bm{X}_\mathrm{rel}=(-97.2\pm23.5,-121.6\pm34.8,-129.8\pm19.0)\;\mathrm{kpc}.
\end{equation}
In order to calculate the relative velocity between the two galaxies in the same reference frame, first we used the values
of the proper motions of M33 from {\cite{m33pm}}, the LOS velocity as used in \cite{vdm12a}, $v_{\mathrm{LOS,
M33}}=-180\pm1\;\mathrm{km\ s}^{-1}$ (\citealt{vdmG}) and applied the correction for the motion of the Sun
$(U_{\odot},V_{\odot}+V_{\mathrm{LSR}},W_{\odot})=(11.1\pm0.7, 255.2\pm5.1,7.25\pm0.36)\;\mathrm{km\ s}^{-1}$
(\citealt{solar1}; \citealt{solar2}). Then we considered the proper motions and LOS velocity of M31 as given by
\cite{vdm12a}, corrected for the same solar values. This yields the first option for the relative velocity
\begin{equation}
	\bm{V}_{\mathrm{rel,vdM12}}=(-23.2\pm34.3,177.4\pm29.2,93.7\pm38.5)\;\mathrm{km\ s}^{-1}.
\end{equation}
The second possible value of the relative velocity was obtained by replacing the values for M31 with those given by
\cite{s16} and correcting for the same solar values. This yields
\begin{equation}
	\bm{V}_{\mathrm{rel,S16}}=(-72.0\pm64.4,86.4\pm48.0,10.6\pm62.1)\;\mathrm{km\ s}^{-1}.
\end{equation}
For all the calculations done here we assumed that error bars follow a Gaussian distribution. We have adopted this approximation in
order to be able to easily transform error bars from one reference frame to another.

\section{B. Orbit integration scheme}
In order to quickly find possible orbits of the two galaxies, without running computationally
costly simulations, we performed semi-analytic orbit integrations, similar to those described in \cite{Patel1},
\cite{dierickx1} or \cite{dierickx2}. We treated both galaxies as two interacting NFW halos (\citealt{nfw}) and we
included dynamical friction from M31 acting on M33. We thus solved the following equations of motion
\begin{equation}       \label{appendixequation}
\begin{aligned}
	\bm{\ddot{x}}_{\mathrm{M33}}=-\nabla \psi_{\mathrm{M31}}+\bm{f}_{\mathrm{DF}},\\
	\bm{\ddot{x}}_{\mathrm{M31}}=-\nabla \psi_{\mathrm{M33}}.
\end{aligned}
\end{equation}
Here, $\bm{x}_{\mathrm{M33}}$ and $\bm{x}_{\mathrm{M31}}$ denote the positions of the M33 and M31 galaxies. The
functions $\psi_{\mathrm{M33}}$ and $\psi_{\mathrm{M31}}$ are NFW potentials of M33 and M31, respectively, given by
\begin{equation}
	\psi_i=-\frac{G M_i}{r[\ln(1+c_i)-c_i/(1+c_i)]}\ln(1+\frac{r}{r_{s,i}}),
\end{equation}
where $M_i$ are the virial masses of the galaxies, $c_i$ are their concentration parameters and $r_{s,i}=r_{v,i}/c_i$
are the scale radii, i.e. the virial radii $r_{v,i}$ divided by concentrations.

For both galaxies we adopted halo parameters that were estimated by modeling the observed rotation curves. For
M33 we assumed $M_\mathrm{M33}=4.38\times10^{11}\;\mathrm{M_{\odot}}$ and $c_{\mathrm{M33}}=11$ (C14). The halo mass
includes the estimate of the mass of baryons in M33, discussed in greater detail in section 3.1. For M31 we
took $M_{\mathrm{M31}}=2\times10^{12}\;\mathrm{M_{\odot}}$ and $c_{\mathrm{M31}}=28$  to fit the rotation curve given
by \cite{m31mass} only with the potential of the halo (see section 3.1). For the supplementary parameters necessary to
compute the NFW profile we adopted the virial overdensity to be $\Delta_c=102$ and the critical density
$\rho_c^0=136\;\mathrm{M_{\odot}}\mathrm{kpc}^{-3}$.

Finally,
$\bm{f}_{\mathrm{DF}}$ in equation (\ref{appendixequation}) denotes the acceleration caused by the dynamical friction,
approximated in our calculations according to the Chandrasekhar formula (\citealt{chandra}):
\begin{equation}
	\bm{f}_{\mathrm{DF}}=-\frac{4 \pi G^2 M_{\mathrm{M33}} \ln \Lambda \rho
	(r)}{v^2}\bigg[\mathrm{erf}(X)-\frac{2X}{\sqrt{\pi}}\mathrm{exp}(-X^2)\bigg] \frac{\bm{v}}{v},
\end{equation}
where $\rho(r)$ is the density of the M31 halo at a given radius $r$, $\bm{v}$ is the relative velocity vector with its
magnitude $|\bm{v}|=v$, and $X=v/\sqrt{2 \sigma}$, where $\sigma$ is the 1D velocity dispersion, which we approximated
according to the formula derived by \cite{zentner}. For the Coulomb logarithm $\ln \Lambda$ we adopted the formula
presented in \cite{hashimoto} and used by \cite{dierickx1} and \cite{dierickx2}:

\begin{equation}
	\ln \Lambda=\ln\bigg(\frac{r}{1.4 \epsilon}\bigg),
\end{equation}

where $\epsilon$ is the softening length of the galaxy subject to dynamical friction. We fit this parameter for M33
by comparing integrated orbits with orbits in preliminary, collisionless, low-resolution simulations. We estimated the
value $\epsilon=28.5$ kpc to be the best match.
To integrate equations (\ref{appendixequation}) we used a symplectic leapfrog integration method as described
in \cite{leapfrog}.

\section{C. Simulations in isolation}

\begin{figure*}
\begin{center}
\includegraphics[width=500pt]{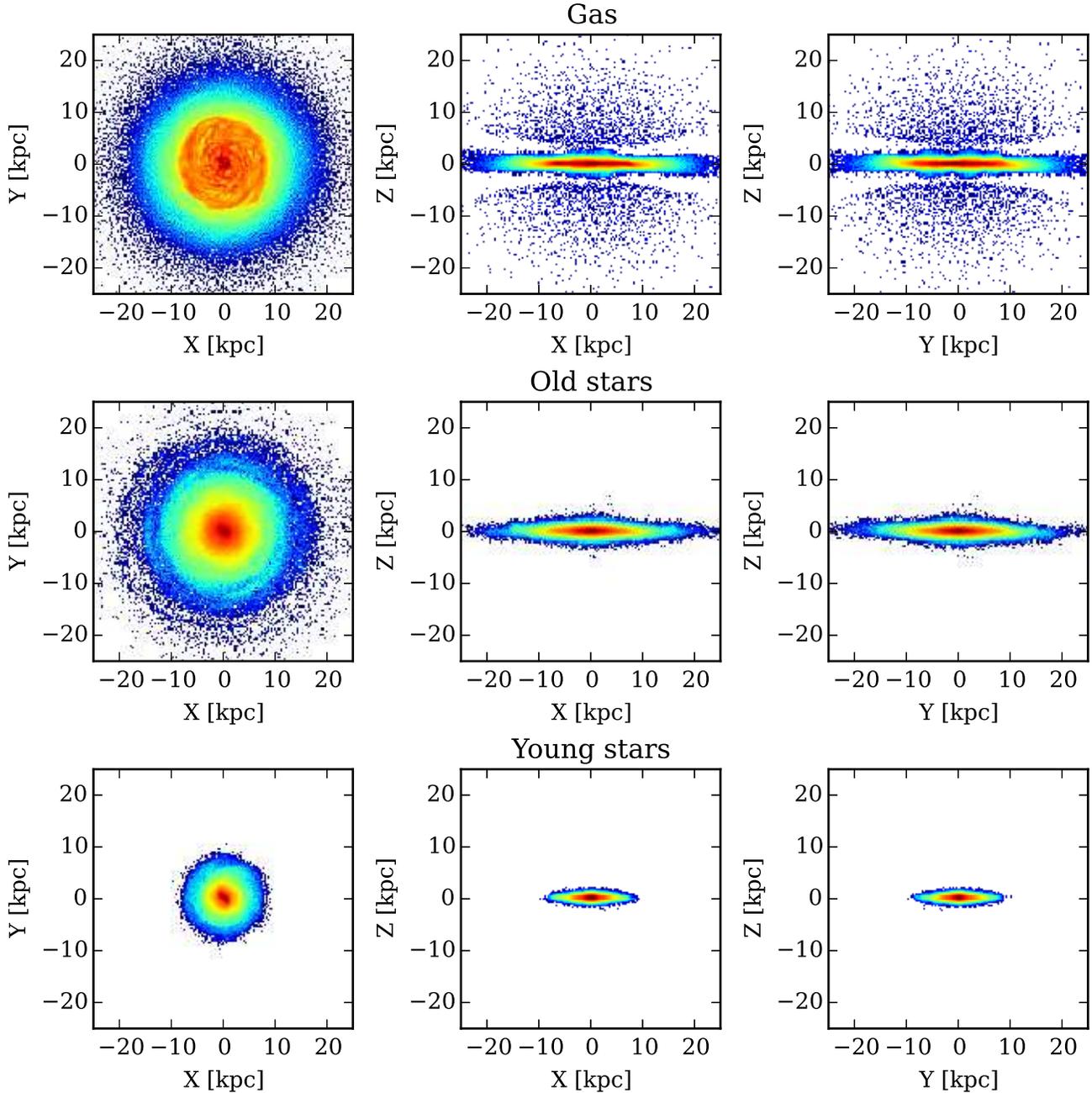}
\caption{Surface density distributions seen from three different directions for the gas, the old and the young stellar
particles for the M33 galaxy simulated in isolation at the time corresponding to the time of the best match for the
fiducial model.}
\label{map_iso}
\end{center}
\end{figure*}

In order to confirm that the morphological features like the gaseous warp, the stellar stream and the grand-design
spiral arms originate from the tidal interaction with M31 and are not induced by secular processes (e.g. feedback or
star formation) in the disk of M33, we evolved this galaxy in isolation for the same time as in the case with the
perturber. Figure~\ref{map_iso} presents surface density maps for all three components of M33 evolved in isolation at
3.45 Gyr, the time of the best match for the fiducial model. A quick look at these images and the comparison with
Figure~\ref{map} is enough to confirm that the aforementioned features are indeed tidally induced, since none of the
components of the isolated galaxy shows signs of similar distortions. Particles present above and below the
gaseous disk in Figure~\ref{map_iso} that are distributed in the hourglass-like shape have been ejected there by feedback
and contribute a very small fraction of the total gaseous content. They are absent in the images of the fiducial
model, most likely due to tidal stripping. The young stellar disk also reveals a small bar which is not a tidal
feature as discussed in section 4.5.

\bibliography{bibliography}

\end{document}